\documentclass{ws-rv9x6}

\usepackage{epsf,amsmath,amssymb,latexsym}

\numberwithin{equation}{section}
\newtheorem{defn}[theorem]{Definition}

\newcommand{\bde}{\begin{defn}}
\newcommand{\ede}{\end{defn}}
\newcommand{\blem}{\begin{lemma}}
\newcommand{\elem}{\end{lemma}}
\newcommand{\bex}{\begin{example}}
\newcommand{\eex}{\end{example}}

\newtheorem{protocol}[theorem]{Protocol}
\newtheorem{observ}[theorem]{Observation}

\newcommand{\bpr}{\begin{proposition}}
\newcommand{\epr}{\end{proposition}}

\newcommand{\CC}{\mathbb{C}}

\newcommand{\Z}{\mathbb{Z}}

\newcommand{\R}{\mathbb{R}}

\newcommand{\SSS}{{\rm SSS}}
\newcommand{\USS}{{\rm USS}}

\begin{document}

\def\sof{\hfill\rule{2mm}{2mm}}
\def\ls{\leq}
\def\gs{\geq}
\def\SS{\mathcal S}
\def\qq{{\bold q}}
\def\txx{{\frac1{2\sqrt{x}}}}

\renewcommand\ss[1]{\sigma_{#1}^{\vrule height5pt width0pt}}
\newcommand\sss[1]{\sigma_{#1}^{-1}}
\newcommand\ssss[1]{\sigma_{#1}^{\pm1}}
\newcommand{\nn}{n}
\newcommand{\NN}{N}
\newcommand\DD[1]{\Delta_{#1}}
\newcommand{\cl}[1]{\overline{\vrule height5pt width0pt#1}}
\newcommand{\ww}{w}
\newcommand\xs{s}
\newcommand\xt{t}
\newcommand\xu{u}
\newcommand\xv{v}
\newcommand\xx{x}

\chapter{Braid Group Cryptography}

\author{David Garber}
\address{Department of Applied Mathematics,  Faculty of Sciences,\\
Holon Institute of Technology, \\52 Golomb Street, PO Box 305,\\ 58102
Holon, Israel \\
{\em E-mail:} garber@hit.ac.il}

\date{\today}

\begin{abstract}
In the last decade, a number of public key cryptosystems based on
combinatorial group theoretic problems in braid groups have been
proposed. We survey these cryptosystems and some known attacks on them.

This survey includes: Basic facts on braid groups and on the Garside
normal form of its elements, some known algorithms for solving
the word problem in the braid group, the major public-key cryptosystems
based on the braid group, and some of the known attacks on these
cryptosystems. We conclude with a discussion of future directions
(which includes also a description of cryptosystems which are based on other
non-commutative groups).
\end{abstract}

\setcounter{tocdepth}{3}

\tableofcontents


\section{Introduction}

In many situations, we need to transfer data in a secure way: credit cards information, health data, security uses, etc. The idea of public-key cryptography in general is to make it possible for two parties to agree on a shared secret key, which they can use to transfer data in a secure way (see \cite{KM}).

There are several known public-key cryptosystems which are based on the {\em discrete logarithm problem}, which is the problem of finding $x$ in the equation $g^x=h$ where $g,h$ are given, and on the {\em factorization problem}, which is the problem of factoring a number to its prime factors: Diffie-Hellman \cite{DH} and RSA \cite{RSA}. These schemes are used in most of the present-day applications using public-key cryptography

\medskip

There are several problems with this situation:
\begin{itemize}
\item {\bf Subexponential attacks on the current cryptosystems' underlying problems:}
    Diffie-Hellman and RSA are breakable in time that is subexponential (i.e. faster than an exponential) in the size of the secret key \cite{AdDe}. The current length of secure keys is at least 1000 bits. Thus, the length of the key should be increased every few years. This makes the encryption and decryption algorithms very heavy.

\item {\bf Quantum computers:} If quantum computers will be implemented in a satisfactory way, RSA will not be secure anymore, since there are polynomial (in $\log(n)$) run-time algorithms of Peter Shor \cite{Shor} which solve the factorization problem and the discrete logarithm problem. Hence, it solves the problems which RSA and Diffie-Hellman are based on (for more information, see for example \cite{Aharonov}).

\item {\bf Too much secure data is transferred in the same method:} It is not healthy that most of the secure data in the world will be transferred in the same method, since in case this method will be broken, too much secure data will be revealed.
\end{itemize}

\medskip

Hence, for solving these problems, one should look for a new public-key cryptosystem which
on one hand will be efficient for implementation and use, and on the  other hand will
be based on a problem which is different from
the discrete logarithm problem and the factorization problem. Moreover, the problem should have no subexponential algorithm for solving it, and it is preferable that it has no known attacks by quantum computers.

\medskip

Combinatorial group theory is a fertile ground for finding hard problems which can serve as a base for a cryptosystem. The braid group defined by Artin \cite{Art} is a very interesting group from many aspects: it has many equivalent presentations in entirely different disciplines; its word problem (to determine whether two elements are equal in the group) is relatively easy to solve, but some other problems (as the conjugacy problem, decomposition problem, and more) seem to be hard to solve.

Based on braid group and its problems, two cryptosystems were suggested about a decade ago: by Anshel, Anshel and Goldfeld in 1999 \cite{AAG} and by Ko, Lee, Cheon, Han, Kang and
Park in 2000 \cite{KLCHKP}. These cryptosystems initiated a wide discussion about the possibilities of cryptography in the braid group especially, and in groups in general.

An interesting point which should be mentioned here is that the conjugacy problem in the braid group attracted people even before the cryptosystems on the braid groups were suggested (see, for example, \cite{ElMo,Gar}). After the cryptosystems were  suggested, some probabilistic solutions were given \cite{GKTTV05,GKTTV06,HS}, but it gave a great push for the efforts to solve the conjugacy problem theoretically in polynomial time (see \cite{BGGM1,BGGM2,BGGM3,Ge,GeGM,GeGM2,GeGM3,LeeLee3a,LeeLee3b,LeeLee4} and many more).

\medskip

The potential use of braid groups in cryptography led to additional
proposals of cryptosystems which are based on apparently hard problems in braid groups (Decomposition problem \cite{SU06}, Triple Decomposition problem \cite{Kurt}, Shifted Conjugacy Search problem \cite{Deh06}, and more) and in other groups, like Thompson Groups \cite{SU05}, polycyclic groups \cite{EiKa} and more. For more information, see the new book of Myasnikov, Shpilrain and Ushakov \cite{MSU08}.

\medskip

In these notes, we try to survey this fascinating subject. Section \ref{sec_braid_group} deals with some different presentations of the braid group. In Section \ref{Sec_normal_forms}, we describe two normal forms for elements in the braid groups. In Section \ref{sec_word_problem}, we give several solutions for the word problem in the braid group. Section \ref{sec_PKC} introduces the notion of public-key cryptography.
In Section \ref{sec_PKC_braid_group}, the first cryptosystems which are based on the braid group are presented. Section \ref{sec_Summit_Sets} is devoted to the theoretical solution to the conjugacy search problem, using the different variants of Summit Sets. In Section \ref{sec_attack_CSP}, we describe some more attacks on the conjugacy search problem.
In Section \ref{sec_new_crypto}, we discuss some more suggestions for cryptosystems based on the braid group and their cryptanalysis. Section \ref{sec_disctributions} deals with the option of changing the distribution for choosing a key. In Section \ref{sec_diff_groups}, we deal with some suggestions for cryptosystems which are based on other non-commutative groups.


\section{The braid group}\label{sec_braid_group}

\subsection{Basic definitions}

The braid groups were introduced by Artin \cite{Art}. There are
several definitions for these groups (see \cite{BiBr,Rol}), and we need
two of them for our purposes.

\subsubsection{Algebraic presentation}

\begin{defn}
For $n \ge 2$, the braid group~$B_n$ is defined by the presentation:
\begin{equation}\label{E:Pres}
\left\langle \ss1, \dots, \ss{n-1} \, \left| \,
\begin{array}{c} \ss i \ss j = \ss j \ss i
\mbox{ for $\vert i - j\vert \ge 2$}\\
\ss i \ss {i+1} \ss i = \ss {i+1} \ss i \ss {i+1} \mbox{ for $\vert
i - j\vert = 1$} \end{array} \right. \right\rangle.
\end{equation}
\end{defn}
This presentation is called the {\it Artin presentation} and the generators
are called {\it Artin's generators}.

An element of $B_n$ will be called an {\it $n$-braid}. For each $n$,
the identity mapping on $\{\ss1, \dots, \ss{n-1}\}$ induces an
embedding of $B_n$ into $B_{n+1}$, so that we can consider an
$\nn$-braid as a particular $(\nn+1)$-braid. Using this, one can define the limit
group $B_\infty$.

Note that $B_2$ is an infinite cyclic group, and hence it is isomorphic to the group~$\Z$
of integers. For $\nn\ge3$, the group~$B_\nn$ is not commutative and
its center is an infinite cyclic subgroup.

When a group is specified using a presentation, each element of the
group is an equivalence class of words with respect to the
congruence generated by the relations of the presentation. Hence,
 every $n$-braid is an
equivalence class of $n$-braid words under the congruence
generated by the relations in Presentation~\eqref{E:Pres}.

\subsubsection{Geometric interpretation}

\begin{figure}[htb]
\epsfysize=3cm \centerline{\epsfbox{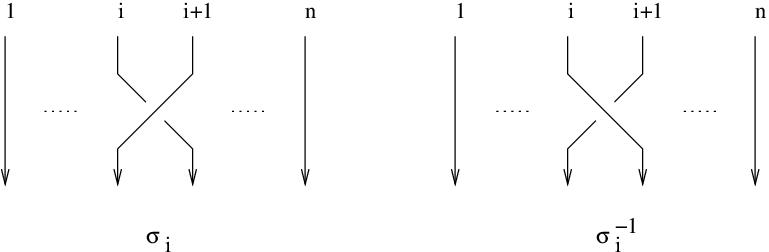}}
\caption{The geometric Artin generators}\label{Artin_gen}
\end{figure}

The elements of $B_n$ can be interpreted as geometric braids with $n$ strands.
One can associate with every braid the planar diagram obtained
by concatenating the elementary diagrams of Figure \ref{Artin_gen}
corresponding to the successive letters.

\begin{figure}[htb]
\epsfysize=4cm \centerline{\epsfbox{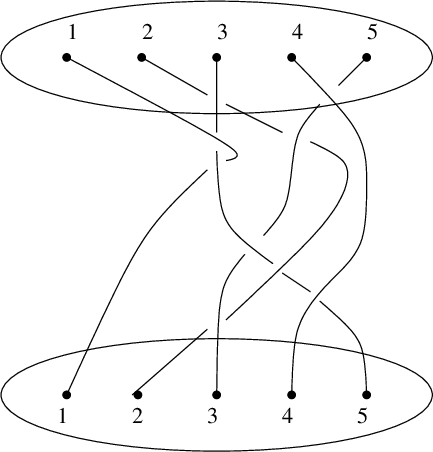}}
\caption{An example of a braid in $B_5$}\label{exam_braid}
\end{figure}

A braid diagram can be seen as induced by a
three-dimensional figure consisting on $n$ disjoint curves
connecting the points $(1, 0, 0), \dots, (n, 0, 0)$ to the points
$(1, 0, 1), \dots, (n, 0, 1)$ in $\R^3$ (see Figure \ref{exam_braid}).

\begin{figure}[htb]
\epsfysize=4cm \centerline{\epsfbox{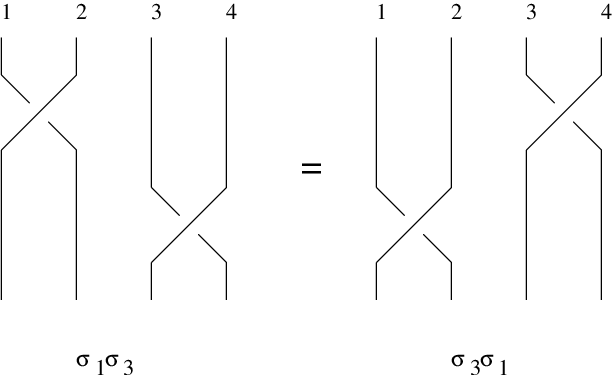}}
\caption{The commutative relation for geometric Artin generators}\label{Artin_comm}
\end{figure}

\begin{figure}[htb]
\epsfysize=5cm \centerline{\epsfbox{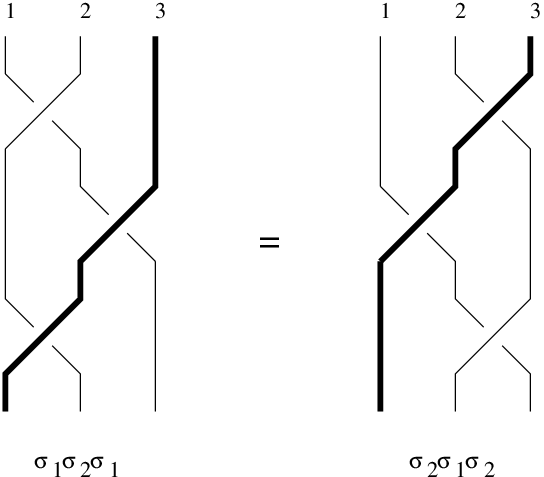}}
\caption{The triple relation for geometric Artin generators}\label{Artin_triple}
\end{figure}

Then the relations
in Presentation~\eqref{E:Pres} correspond to ambient isotopy, that is: to
continuously move the curves without moving their ends and without
allowing them to intersect (see Figures
\ref{Artin_comm} and \ref{Artin_triple}); the converse implication,
i.e., the fact that the projections of isotopic 3-dimensional figures can always
be encoded in words connected by presentation~\eqref{E:Pres} was proved by Artin
in \cite{Art}. Hence, the word problem in the braid group for the
Presentation~\eqref{E:Pres} is also the {\it braid isotopy problem},
and thus it is closely related to the much more difficult knot
isotopy problem.

\subsection{Birman-Ko-Lee presentation} 
\begin{figure}[htb]
\epsfysize=4cm \centerline{\epsfbox{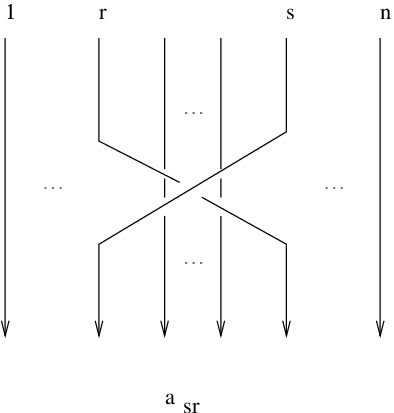}}
\caption{The band generator}\label{BKL_gen}
\end{figure}

Like Artin's generators,
the generators of Birman-Ko-Lee \cite{BKL98} are braids in which
exactly one pair of strands crosses. The difference is that Birman-Ko-Lee's generators
includes arbitrary transpositions of strands $(i, j)$
instead of adjacent transpositions $(i, i + 1)$ in the Artin's
generators. For each $t,s$ with $1 \leq s < t \leq n$, define
the following element of $B_n$:
$$a_{ts} = (\sigma_{t-1}\sigma_{t-2} \cdots \sigma_{s+1})\sigma_s (\sigma_{s+1}^{-1}\cdots \sigma_{t-2}^{-1} \sigma_{t-1}^{-1})$$
See Figure \ref{BKL_gen} for an example (note that the braid $a_{ts}$ is an
elementary interchange of the $t$th and $s$th strands, with all
other strands held fixed, and with the convention that the strands
being interchanged pass in front of all intervening strands). Such an
element is called a {\it band generator}.

Note that the usual Artin generator $\sigma_t$ is the band generator
$a_{t+1,t}$.

This set of generators satisfies the following relations (see
\cite[Proposition 2.1]{BKL98} for a proof):
\begin{itemize}
\item $a_{ts} a_{rq} = a_{rq} a_{ts}$ if $[s,t]\cap [q,r]=\emptyset$.
\item $a_{ts} a_{sr} = a_{tr} a_{ts} = a_{sr} a_{tr}$ for $1 \leq r < s < t \leq n$.
\end{itemize}

\begin{figure}[htb]
\epsfysize=4cm \centerline{\epsfbox{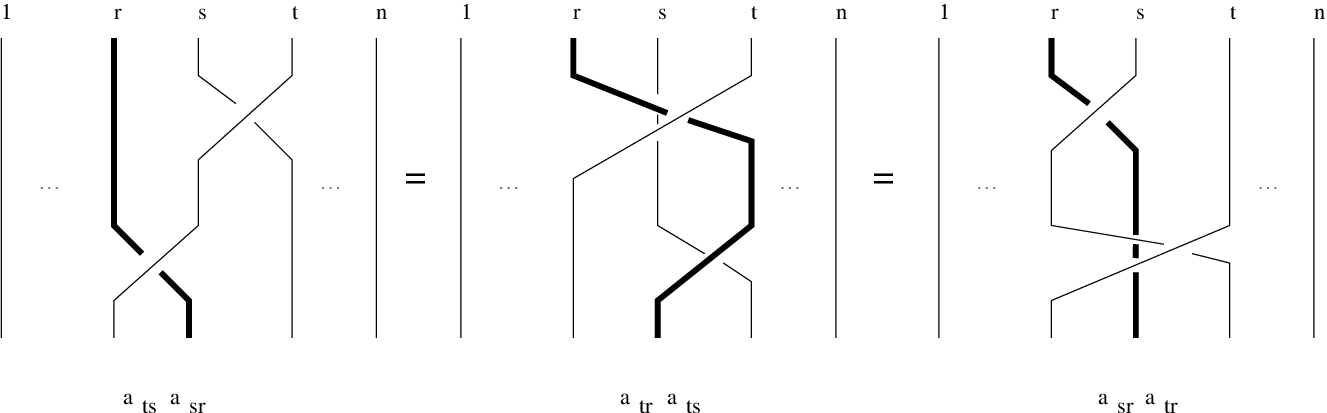}}
\caption{The second relation of the Birman-Ko-Lee presentation}\label{BKL_comm}
\end{figure}

For a geometric interpretation of the second relation, see Figure \ref{BKL_comm}.

\subsubsection{A geometric viewpoint on the difference between presentations}\label{geom-viewpoint-sec}

A different viewpoint on the relation between the two presentations is as follows: one can think
on the braid group as the isotopy classes of boundary-fixing
homeomorphisms on the closed disk $D_n \subset \CC^2$ centered at $0$  with $n$ punctures \cite{Art}.

\begin{figure}[!ht]
\epsfysize=4cm \centerline{\epsfbox{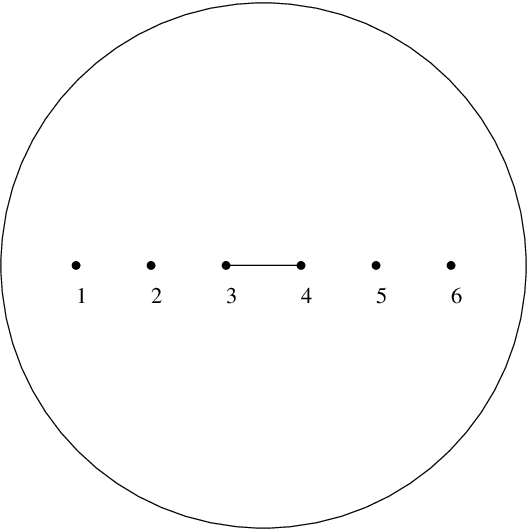}}
\caption{The Artin generator $\sigma_3$}\label{pres_artin_gen}
\end{figure}

In this viewpoint, for presenting the Artin generators, we locate the punctures on the real line,
and the generator $\sigma_i$ is the homeomorphism which exchanges the points $i$ and $i+1$ along the
real line (see Figure \ref{pres_artin_gen}).

\begin{figure}[!ht]
\epsfysize=4cm \centerline{\epsfbox{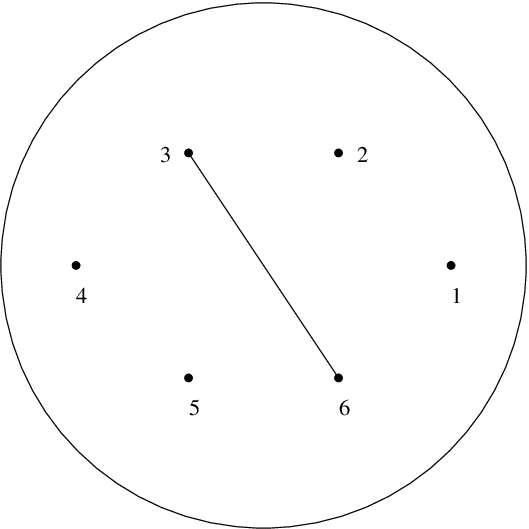}}
\caption{The Birman-Ko-Lee generator $a_{63}$}\label{pres_BKL_gen}
\end{figure}

On the other hand, for illustrating the generators $a_{ts}$ of the Birman-Ko-Lee presentation,
let us take the punctures organized as the vertices of a $n$-gon contained in the disk $D_n$.
Now, the generator $a_{ts}$ is the homeomorphism which exchanges the points $t$ and $s$ along the
chord connecting them (see Figure \ref{pres_BKL_gen}).

For more information, see  \cite{BDM,Bra}.


\section{Normal forms of elements in the braid group}\label{Sec_normal_forms}

A {\it normal form} of an element in a group is a unique presentation
to each element in the group.

Having a normal form for elements in the group is very useful,
since it lets us compare two elements, so it gives
a solution for the word problem:
\begin{problem}
Given a braid $w$, does $w\equiv\varepsilon$ hold, i.e., does $w$
represent the unit braid $\varepsilon$ (see Figure \ref{unit_braid})?

\begin{figure}[!htf]
\epsfysize=4cm \centerline{\epsfbox{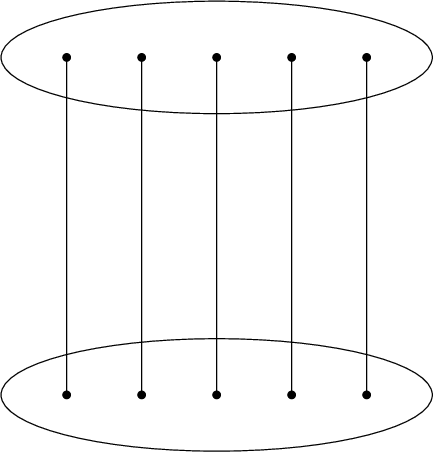}}
\caption{The unit braid $\varepsilon\in B_5$ }\label{unit_braid}
\end{figure}
\end{problem}

Since $B_n$ is a group, the above problem is equivalent to the following
problem:
\begin{problem}
Given two braids~$\ww, \ww'$, does $\ww\equiv\ww'$ hold, i.e.,
do $\ww$ and $\ww'$ represent the same braid?
\end{problem}
Indeed, $w\equiv\ww'$ is equivalent to $w^{-1}w'\equiv\varepsilon$,
where $w^{-1}$ is the word obtained from $w$ by reversing the
order of the letters and exchanging~$\sigma_i$ and~$\sigma^{-1}_i$
everywhere.

Also, the normal form gives a canonical representative of each equivalence class.

\medskip

We present here two known normal forms of elements in the braid
group. For more normal forms, see \cite{Bre,Deh07a,DyWi}.

\subsection{Garside normal form}\label{GNF_Sec}

The Garside normal form is initiated in the work of Garside \cite{Gar},
and several variants have been described in
several partly independent papers \cite{Adj,Del,ElMo,ECHLPT,Thu}.

\medskip

We start by defining a {\it positive braid} which is a braid which can be
written as a product of positive powers of Artin generators. We
denote the set of positive braids by $B_n^+$. This set has a
structure of a monoid under the operation of braid concatenation.

 An important example of a positive braid, which has a
central role in the Garside normal form, is the {\it fundamental braid}
$\Delta_n \in B_n$:
$$\Delta_n = (\sigma_1 \cdots \sigma_{n-1})(\sigma_1 \cdots \sigma_{n-2}) \cdots \sigma_1$$

\begin{figure}[htb]
\epsfysize=4cm \centerline{\epsfbox{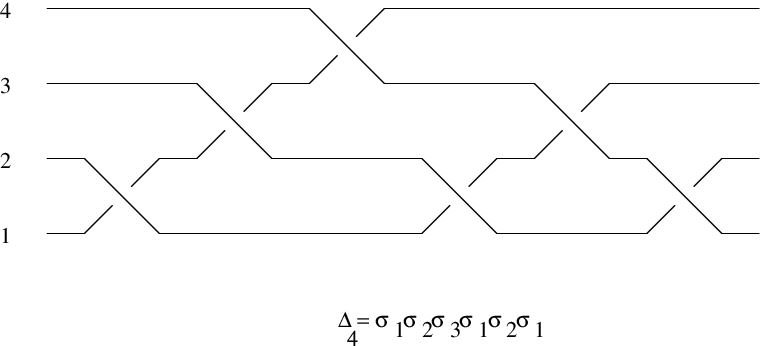}}
\caption{The fundamental braid $\Delta_4$}\label{Artin_delta}
\end{figure}

Geometrically, $\Delta_n$ is the braid on $n$ strands, where any two
strands cross positively exactly once (see Figure \ref{Artin_delta}).

The fundamental braid has several important properties:
\begin{enumerate}
\item For any generator $\sigma_i$, we can write $\Delta_n= \sigma_i A = B \sigma_i$ where
$A,B$ are positive braids.
\item For any generator $\sigma_i$, the following holds:
$$\tau(\sigma_i) = \Delta_n^{-1} \sigma_i \Delta_n = \sigma_{n-i}$$
(the inner automorphism $\tau$ on $B_n$ is called the {\em shift map}).
\item $\Delta_n^2$ is the generator of the center of $B_n$.
\end{enumerate}

Now, we introduce {\it permutation braids}. One can define a partial
order on the elements of $B_n$: for $A,B \in B_n$, we say that $A$ is a {\em prefix} of $B$
and write $A \preceq B$ if $B = AC$ for some $C$ in $B^+_n$. Its simple
properties are:
\begin{enumerate}
\item $B \in B^+_n \Leftrightarrow  \varepsilon \preceq B$
\item $A \preceq B \Leftrightarrow B^{-1} \preceq A^{-1}$.
\end{enumerate}

$P \in B_n$ is a {\it permutation braid} (or a {\it simple braid})
if it satisfies: $\varepsilon \preceq P \preceq \Delta_n$. Its name comes
from the fact that there is a bijection between the set of
permutation braids in $B_n$ and the symmetric group $S_n$ (there is
a natural surjective map from $B_n$ to $S_n$ defined by sending $i$
to the ending place of the strand which starts at position $i$, and if we
restrict ourselves to permutation braids, this map is a bijection).
Hence, we have $n!$ permutation braids.

Geometrically, a permutation braid is a braid on $n$ strands, where
any two strands cross positively {\it at most} once.

Given a permutation braid $P$, one can define a {\it starting set} $S(P)$
and a {\it finishing set} $F(P)$ as follows:
$$S(P)=\{ i | P=\sigma_i P' \mbox{ for some } P'\in B_n^+ \}$$
$$F(P)=\{ i | P=P' \sigma_i  \mbox{ for some } P'\in B_n^+ \}$$
The starting set is the indices of the generators which can start a
presentation of $P$. The finishing set is defined similarly.
For example, $S(\Delta_n)=F(\Delta_n)=\{1,\dots ,n-1\}$.

A {\it left-weighted decomposition} of a positive braid $A \in
B_n^+$ into a sequence of permutation braids is:
$$A = P_1 P_2 \cdots P_k$$
where $P_i$ are permutation braids, and $S(P_{i+1}) \subset F(P_i)$,
i.e.\ any addition of a generator from $P_{i+1}$ to $P_i$, will
convert $P_i$ into a braid which is not a permutation braid.

\begin{example}
The following braid is left-weighted:

\epsfysize=4cm \centerline{\epsfbox{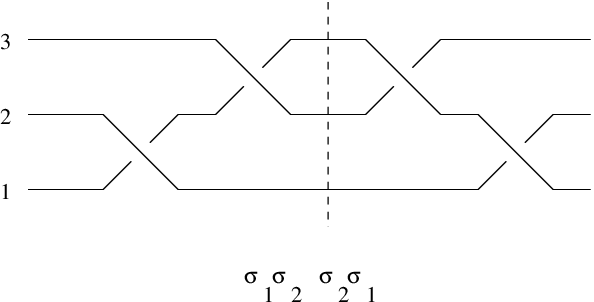}}

\medskip

The following braid is not left-weighted, due to the circled crossing which can be moved to
the first permutation braid:

\epsfysize=4cm \centerline{\epsfbox{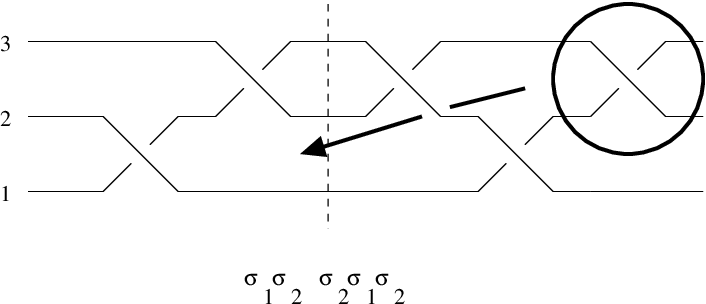}}

Now, we show it algebraically:
$$\sigma_1\sigma_2 \cdot \sigma_2\sigma_1\sigma_2 = \sigma_1\sigma_2 \cdot \sigma_1\sigma_2\sigma_1=\sigma_1\sigma_2\sigma_1\cdot\sigma_2\sigma_1$$
\end{example}

\medskip

The following theorem introduces the {\it Garside normal form} (or
{\it left normal form} or {\it greedy normal form}) and states its uniqueness:

\begin{theorem}\label{GNF}
For every braid $w \in B_n$, there is a unique presentation given
by:
$$w= \Delta_n^r P_1 P_2 \cdots P_k$$
where $r \in \mathbb{Z}$ is maximal, $P_i$ are permutation braids, $P_k\neq \varepsilon$ and $P_1 P_2
\cdots P_k$ is a left-weighted decomposition.
\end{theorem}

For converting a given braid $w$ into its Garside normal form we have
to perform the following steps:
\begin{enumerate}
\item For any negative power of a generator, replace $\sigma_i^{-1}$ by
$\Delta_n^{-1} B_i$ where $B_i$ is a permutation braid.
\item Move any appearance of $\Delta_n$ to the left using the
relation:\break $\Delta_n^{-1} \sigma_{i} \Delta_n = \tau (\sigma_i) = \sigma_{n-i}$. So we get:
$w=\Delta_n^{r'} A$ where $A$ is a positive braid.
\item Write $A$ as a left-weighted decomposition of permutation
braids. The way to do this is as follows: Take $A$, and break it into permutation braids
(i.e.\ we take the longest possible sequences of generators which are still permutation braids).
Then we get: $A=Q_1 Q_2 \cdots Q_j$ where each $Q_i$ is a permutation braid. For each $i$, we compute
the finishing set $F(Q_i)$ and the starting set $S(Q_{i+1})$. In case the starting set is not contained in the finishing set, we take a generator $\sigma \in S(Q_{i+1}) \setminus F(Q_i)$, and using the relations of the braid group we move it from $Q_{i+1}$ to $Q_i$. Then, we get the decomposition
$A=Q_1 Q_2 \cdots Q'_i Q'_{i+1} \cdots Q_j$. We continue this process till we have
$S(Q_{i+1}) \subseteq F(Q_i)$ for every $i$, and then we have a left-weighted decomposition as needed.
For more details, see \cite{ElMo} and \cite[Proposition 4.2]{GeGM3} (in the latter reference, it is done based on their new idea of local slidings, see Section \ref{subsec_sliding} below) .
\end{enumerate}

\begin{example}
Let us present the braid  $w=\sigma_1 \sigma_3^{-1} \sigma_2 \in B_4$ in
Garside normal form. First, we should replace $\sigma_3^{-1}$ by:
$\Delta_4^{-1}\sigma_3\sigma_2\sigma_1\sigma_3\sigma_2$, so we get:
$$w=\sigma_1 \cdot \Delta_4^{-1}\sigma_3\sigma_2\sigma_1\sigma_3\sigma_2 \cdot \sigma_2$$
Now, moving $\Delta_4$ to the left yields:
$$w=\Delta_4^{-1} \cdot \sigma_3 \sigma_3\sigma_2\sigma_1\sigma_3\sigma_2 \sigma_2$$
Decomposing the positive part into a left-weighted decomposition, we
get:
$$w = \Delta_4^{-1} \cdot \sigma_2 \sigma_1 \sigma_3 \sigma_2 \sigma_1 \cdot \sigma_1 \sigma_2$$
\end{example}

The complexity of transforming a word into a canonical form with
respect to the Artin presentation is $O(|W|^2 n \log n)$ where
$|W|$ is the length of the word in $B_n$ \cite[Section 9.5]{ECHLPT}.

\medskip

In a similar way, one can define a {\it right normal form}. A {\it
right-weighted decomposition} of a positive braid $A \in B_n^+$ into
a sequence of permutation braids is:
$$A = P_k \cdots P_2  P_1$$
where $P_i$ are permutation braids, and $F(P_{i+1}) \subset S(P_i)$,
i.e.\ any addition of a generator from $P_{i+1}$ to $P_i$, will
convert $P_i$ into a braid which is not a permutation braid.

Now, one has the following theorem about the {\it right normal
form} and its uniqueness:

\begin{theorem}\label{RNF}
For every braid $w \in B_n$, there is a unique presentation given
by:
$$w= P_k \cdots P_2 P_1 \Delta_n^r$$
where $r \in \mathbb{Z}$, $P_i$ are permutation braids, and $P_k
\cdots P_2 P_1$ is a right-weighted decomposition.
\end{theorem}

For converting a given braid $w$ into its right normal form we have
to follow three steps, similar to those of the Garside normal form:
We first replace $\sigma_i^{-1}$ by $B_i \Delta_n^{-1}$. Then, we
move any appearance of $\Delta_n$ to the right side. Then, we get:
$w=A \Delta_n^{r'}$ where $A$ is a positive braid. The last step is
to write $A$ as a right-weighted decomposition of permutation
braids.

\medskip

Now we define the {\it infimum} and the {\it supremum} of a braid $w$: For $w
\in B_n$, set ${\rm inf}(w) = \max\{r : \Delta_n^r \preceq w \}$ and
${\rm sup}(w) = \min\{s : w \preceq \Delta_n^s\}$.

One can easily see that if $w = \Delta_n^m P_1 P_2 \cdots P_k$ is
the Garside normal form of $w$, then: ${\rm inf}(w)=m,\  {\rm
sup}(w)=m+k$.

The {\it canonical length of $w$} (or {\it complexity of $w$}),
denoted by $\ell(w)$, is given by ${\rm len}(w) = {\rm sup}(w)-
{\rm inf}(w)$. Hence, if $w$ is given in its normal form, the
canonical length is the number of permutation braids in the form.

\subsection{Birman-Ko-Lee canonical form}

Based on the presentation of Birman, Ko and Lee \cite{BKL98}, they
give a new canonical form for elements in the braid group.

\begin{figure}[htb]
\epsfysize=4cm \centerline{\epsfbox{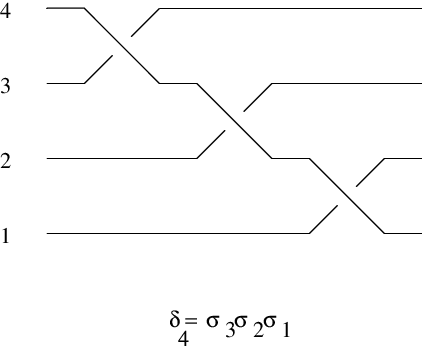}}
\caption{The fundamental braid $\delta_4$}\label{BKL_delta}
\end{figure}

They define a new fundamental word:
$$\delta_n = a_{n,n-1} a_{n-1,n-2} \cdots a_{2,1}= \sigma_{n-1} \sigma_{n-2} \cdots \sigma_1$$
See Figure \ref{BKL_delta} for an example for $n=4$.

One can easily see the connection between the new fundamental word
and Garside's fundamental word $\Delta_n$:
$$\Delta_n^2=\delta_n^n$$

The new fundamental word $\delta_n$ has important properties,
similar to $\Delta_n$:
\begin{enumerate}
\item For any generator $a_{sr}$, we can write $\delta_n= a_{sr} A = B a_{sr}$ where
$A,B$ are positive braids (with respect to the Birman-Ko-Lee generators)
\item For any generator $a_{sr}$, the following holds:
$a_{sr} \delta_n = \delta_n a_{s+1,r+1}$.
\end{enumerate}

\medskip

Similar to Garside's normal form of braids, each element of $B_n$
has the following unique form in terms of the band generators:
$$w = \delta_n^j A_1 A_2 \cdots A_k,$$
where $A = A_1 A_2 \cdots A_k$ is positive, $j$ is maximal and
$k$ is minimal for all such representations, also the $A_i$'s are
positive braids which are determined uniquely by their associated
permutations (see \cite[Lemma 3.1]{BKL98}). Note that
not every permutation corresponds to a canonical factor. We will refer to
Garside's braids $P_i$ as {\it permutation braids}, and to the Birman-Ko-Lee
braids $A_i$ as {\it canonical factors}.

Note that there are $C_n = \frac{(2n)!}{ n!(n + 1)!}$ (the  $n$th
Catalan number) different canonical factors for the band-generators
presentation \cite[Corollary 3.5]{BKL98}, whence there are $n!$
different permutation braids for the Artin presentation. Since $C_n$
is much smaller than $n!$, it is sometimes computationally easier to
work with the band-generators presentation than the Artin
presentation (see also Section \ref{length_function} below).

As  in Garside's normal form, there is an algorithmic way to convert
any braid to this canonical form: we first convert any negative
power of a generator to $\delta_n^{-1} A$ where $A$ is positive.
Then, we move all the $\delta_n$ to the left, and finally we
organize the positive word in a left-weighted decomposition of
canonical factors.

The complexity of transforming a word into a canonical form with
respect to the Birman-Ko-Lee presentation is $O(|W|^2 n)$, where
$|W|$ is the length of the word in $B_n$ \cite{BKL98}.

As in Garside's normal form, one can define infimum, supremum and
canonical length for the canonical form of the Birman-Ko-Lee
presentation.


\section{Algorithms for solving the word problem in braid group}\label{sec_word_problem}

Using $\varepsilon$ for the unit word (see Figure \ref{unit_braid}), the {\it word problem} is the
following algorithmic problem:
\begin{problem}
Given one braid word $w$, does $w\equiv\varepsilon$ hold, i.e., does $w$
represent the unit braid~$\varepsilon$?
\end{problem}

In this section, we will concentrate on some solutions for the
word problem in the braid group.

\subsection{Dehornoy's handles reduction}
The process of {\it handle reduction} has been introduced by Dehornoy
\cite{Deh97}, and one can see it as an extension of the free
reduction process for free groups. Free reduction consists
of iteratively deleting all patterns of the form $xx^{-1}$ or
$x^{-1}x$: starting with an arbitrary word $w$ of length $m$, and
no matter on how the reductions are performed, one finishes in at
most $m/2$ steps with a unique reduced word, i.e., a word that
contains no $xx^{-1}$ or $x^{-1}x$.

Free reduction is possible for any group presentation, and
in particular for $B_n$, but it does not solve the word problem:
there exist words that represent $\varepsilon \in B_n$, but do not freely
reduce to the unit word. For example, the word $\sigma_1 \sigma_2
\sigma_1 \sigma_2^{-1} \sigma_1^{-1} \sigma_2^{-1}$ represents the
unit word, but free reductions can not reduce it any more.

The handle reduction process generalizes free reduction and involves
not only patterns of the form $xx^{-1}$ or $x^{-1} x$, but also more
general patterns of the form $\sigma_i \cdots \sigma_i^{-1}$ or
$\sigma^{-1}_i \cdots \sigma_i$:

\begin{defn}
{\em A $\sigma_i$-handle} is a braid word of the form
$$w = \sigma_i^e w_0 \sigma_{i+1}^d w_1 \sigma_{i+1}^d \cdots \sigma_{i+1}^d w_m \sigma_i^{-e},$$
with $e, d = \pm 1, m \geq 0$, and $w_0,\dots , w_m$ containing no
$\sigma_j^{\pm 1}$ with $j \leq i + 1$.

The {\em reduction of $w$} is defined as follows:
$$ w' = w_0 \sigma^{-e}_{i+1} \sigma_i^d \sigma^e_{i+1} w_1 \sigma^{-e}_{i+1} \sigma_i^d \sigma^e_{i+1}
\cdots \sigma^{-e}_{i+1} \sigma_i^d \sigma^e_{i+1} w_m,$$ i.e., we
delete the initial and final letters $\sigma_i^{\pm 1}$, and we
replace each letter $\sigma_{i+1}^{\pm 1}$ with $\sigma^{-e}_{i+1}
\sigma_i^{\pm 1} \sigma^e_{i+1}$ (see Figure \ref{handle_reduct}, taken from \cite{Deh04}).

\begin{figure}[htb]
\epsfysize=3cm \centerline{\epsfbox{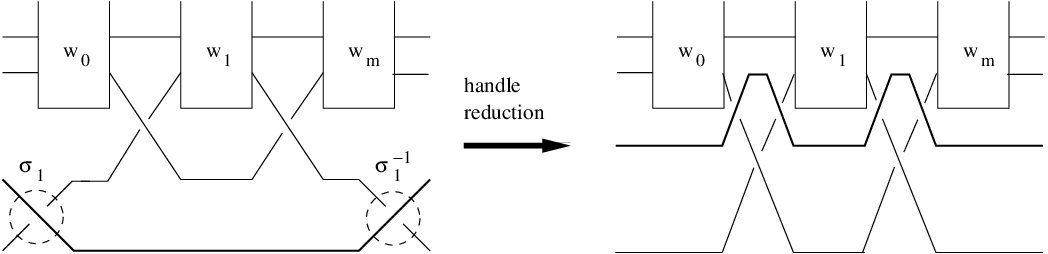}}
\caption{An example for a handle reduction (for $\sigma_1$). The two circled crossings in the left side are the start and the end of the handle}\label{handle_reduct}
\end{figure}

\end{defn}

Note that a braid of the form $\sigma_i \sigma_i^{-1}$ or
$\sigma_i^{-1} \sigma_i$ is a handle, and hence we see that the handle reduction process
generalizes the free reduction process.

Reducing a braid yields an equivalent braid: as illustrated in
Figure \ref{handle_reduct}, the $(i + 1)$th strand in a $\sigma_i$-handle forms a
sort of handle, and the reduction consists of pushing that strand so
that it passes above the next crossings instead of below. So, as in
the case of a free reduction, if there is a reduction sequence from a
braid $w$ to $\varepsilon$, i.e., a sequence $w
= w_0 \to w_1 \to \cdots \to w_N = \varepsilon$ such that, for each $k$,
$w_{k+1}$ is obtained from $w_k$ by replacing some handle of $w_k$
by its reduction, then $w$ is equivalent to $\varepsilon$, i.e., it
represents the unit word $\varepsilon$.

The following  result of Dehornoy \cite{Deh97} shows the converse
implication and the termination of the process of handle reductions:

\begin{prop}
Assume that $w \in B_n$ has a length $m$. Then every reduction
sequence starting from $w$ leads in at most $2^{m^4 n}$ steps to
an irreducible braid (with respect to Dehornoy's reductions).
Moreover, the unit word $\varepsilon$ is the only irreducible word
in its equivalence class, hence $w$ represents the unit braid if
and only if  any reduction sequence starting from $w$ finishes with
the unit word.
\end{prop}

A braid may contain many handles, so building an actual
algorithm requires to fix a strategy prescribing in which order the
handles will be reduced. Several variants have been considered; as
can be expected, the most efficient ones use a ``Divide and Conquer''
trick.

For our current purpose, the important fact is that, although the
proved complexity upper bound of the above proposition is very high,
handle reduction is extremely efficient in practice, even more than
the reduction to a normal form, see \cite{Deh04}.

\begin{remark}
In \cite{Deh07c}, Dehornoy gives an alternative proof for the
convergence of the handle reduction algorithm of braids which is
both more simple and more precise than the one in his original paper
on handle reductions \cite{Deh97}.
\end{remark}

\subsection{Action on the fundamental group}

As we have pointed out at Section \ref{geom-viewpoint-sec}, the braid group can be thought of as the isotopy classes of boundary-fixing homeomorphisms on the closed disk $D_n \subset \CC^2$ centered at $0$  with $n$ punctures $p_1, \dots,p_n$ \cite{Art}. It means that two elements are the same if their actions on $\pi _1(D_n \setminus \{ p_1, \dots, p_n \},u)$ are equal.

\begin{figure}[htb]
\epsfysize=4cm \centerline{\epsfbox{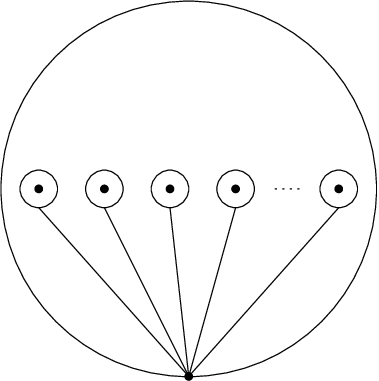}}
\caption{A geometric base}\label{geom_base}
\end{figure}

In \cite{GKT}, we propose the following solution for the word problem: we start with a geometric base for  $\pi _1(D_n \setminus \{ p_1, \dots, p_n \},u)$ presented in Figure
\ref{geom_base}.

Now, we apply the two braids on this initial geometric base. If the resulting bases are the same up to isotopy, it means that the braids are equal, otherwise they are different.

\begin{figure}[htb]
\epsfxsize=12cm \centerline{\epsfbox{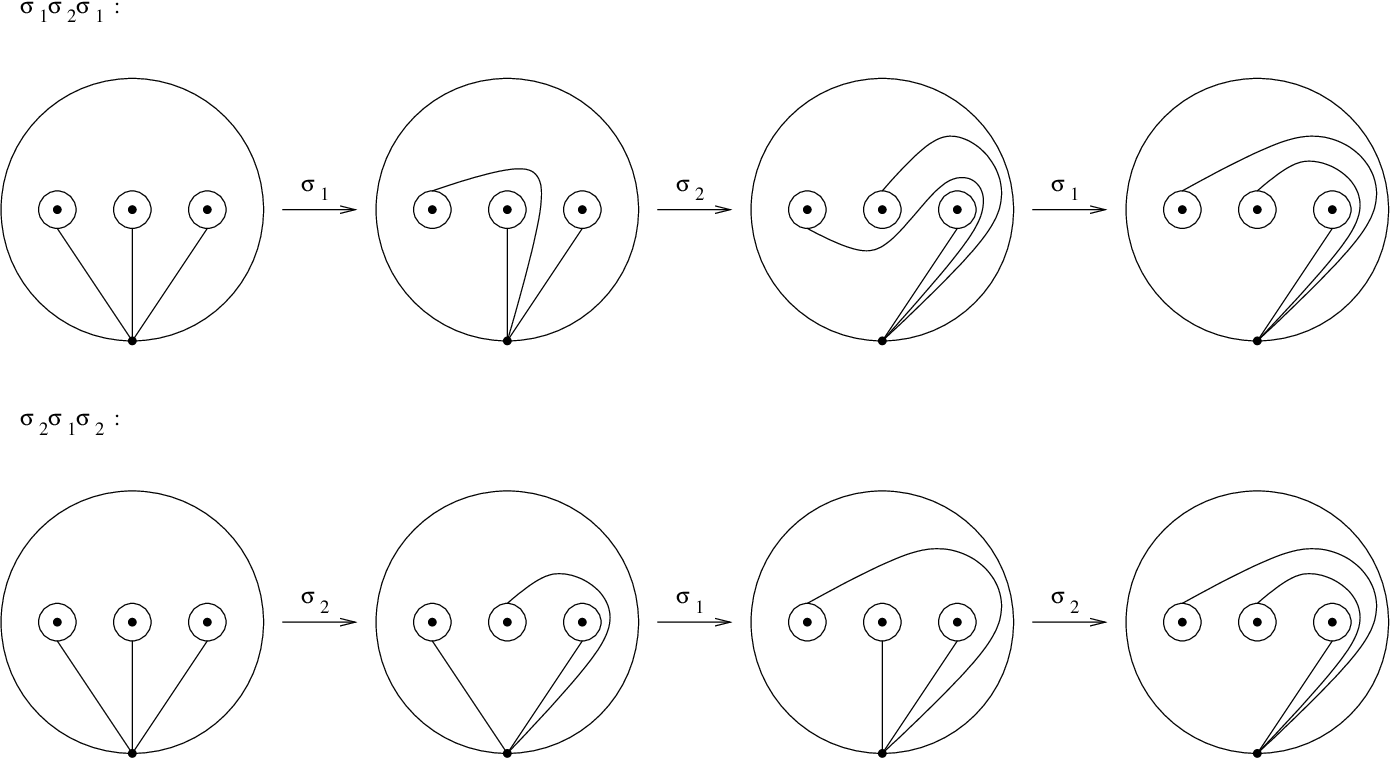}}
\caption{An example of applications of two equal braids $\sigma_1 \sigma_2 \sigma_1 = \sigma_2 \sigma_1 \sigma_2$ on the initial geometric base}\label{exam_geom_base}
\end{figure}

In Figure \ref{exam_geom_base}, there is a simple example of two equal braids which result the same base.

This algorithm is very quick and efficient for short words, but its worst case is exponential. For more details on its implementation, see \cite{GKT}.

\medskip

For more solutions for the word problem for the braid groups, see \cite{Dy}.



\section{What is Public-Key Cryptography?}\label{sec_PKC}
The idea of Public-Key Cryptography (PKC) was invented by Diffie and Hellman \cite{DH}.
At the heart of this concept is the idea of using a one-way
function for encryption (see the survey paper of Koblitz and Menezes \cite{KM}).

The functions used for encryption belong to a special class of {\it
one-way functions} that remain one-way only if some information (the
decryption key) is kept secret. If we use informal terminology, we
can define a {\it public-key encryption function} as a map from
plain text message units to ciphertext message units that can be
feasibly computed by anyone having the public key, but whose inverse
function (which deciphers the ciphertext message units) cannot be
computed in a reasonable amount of time without some additional
information, called the {\it private key}.

This means that everyone can send a message to a given person using the
same enciphering key, which can simply be looked up in a public directory whose
contents can be authenticated by some means. There is no need for the sender
to have made any secret arrangement with the recipient; indeed, the recipient
need never have had any prior contact with the sender at all.

\medskip

Some of the purposes for which public-key cryptography has been applied
are:

\begin{itemize}

\item {\bf Confidential message transmission:} Two people want to exchange messages in the
open airwaves, in such a way that an intruder observing
the communication cannot understand the messages.

\item {\bf Key exchange} or {\bf Key agreement}: Two people using the open airwaves want to agree
upon a secret key for use in some symmetric-key cryptosystem. The agreement should
be in such a way that an intruder observing
the communication cannot deduce any useful information about their shared secret.

\item {\bf Authentication:} The prover wishes to convince the verifier that he knows
the private key without enabling an intruder watching the communication to
deduce anything about his private key.

\item {\bf Signature:} The target in this part is:
The sender of the message has to send the receiver a (clear or
ciphered) message together with a signature proving the origin of
the message. Each signature scheme may lead to an authentication
scheme: in order to authenticate the sender, the receiver can send a
message to the sender, and require that the sender signs this
message.
\end{itemize}

Now, we give some examples of the most famous and well-known public-key
cryptosystems.

\subsection{Diffie-Hellman}
In 1976, Diffie and Hellman \cite{DH} introduced a key-exchange
protocol which is based on the apparent difficulty of computing
logarithms over a finite field ${\mathbb F}_q$ with $q$ elements and
on some commutative property of the exponent.

\medskip

Their key-exchange protocol works as follows:

\begin{protocol}\

{\em Public keys:} $q$ and a primitive element $\alpha$.

{\em Private keys:} Alice: $X_i$; Bob: $X_j$.

\medskip

{\em Alice:} Sends Bob $Y_i= \alpha^{X_i} \pmod q$.

{\em Bob:} Sends Alice $Y_j= \alpha^{X_j} \pmod q$

\medskip

{\em Shared secret key:} $K_{ij} = \alpha^{X_i X_j} \pmod q$
\end{protocol}

$K_{ij}$ is indeed a shared key since Alice can compute  $K_{ij} =
Y_j ^{X_i} \pmod q$ and Bob can compute  $K_{ij} = Y_i ^{X_j} \pmod
q$.

\medskip

This method is secured due to the hardness of the Discrete Logarithm Problem.

\subsection{RSA}

Rivest, Shamir and Adleman \cite{RSA} introduced one of the most
famous and common
cryptosystem, which is called RSA. This method is widely used in commerce.

Find two large prime numbers $p$ and $q$, each about 100 decimal digits long.
Let $n = pq$ and $\phi = \phi(n)=(p - 1)(q - 1)$ (the Euler number).
Choose a random integer $E$ between $3$ and $\phi$ that has no common factors
with $\phi$. It is easy to find an integer $D$ that is the "inverse" of $E$ modulo
$\phi$, that is, $D \cdot E$ differs from $1$ by a multiple of $\phi$.

Alice makes $E$ and $n$ public. All the other quantities here are
kept secret.

The encryption is done as follows: Bob, who wants to send a plain
text message $P$ to Alice, that is an integer between $0$ and $n -
1$, computes the ciphertext integer $C = P ^E \pmod n$. (In other
words, raise $P$ to the power $E$, divide the result by $n$, and
$C$ is the remainder). Then, Bob sends $C$ to Alice.

For decrypting  the message, Alice uses the secret decryption number
$D$ for finding the plain text $P$ by computing: $P = C^D \pmod n$.

\medskip

This method is  currently secure, since in order to determine the
secret decryption key $D$ (for decrypting the message), the intruder
should factor the 200 or so digits number $n$, which is a very hard
task.


\section{First cryptosystems which are based on the braid groups}\label{sec_PKC_braid_group}

In this section, we describe first cryptosystems which are based on the braid groups. We start with the
definition of some apparently hard problems which the cryptosystems are based on. After that, we describe
first two key-exchange protocols which are based on the braid group. We finish the section with some more cryptosystems based on the braid group.

\subsection{Underlying problems for cryptosystems in the braid group}

We list here several apparently hard problems in the braid group, which are the base of many  cryptosystems in the braid group:

\medskip

\begin{itemize}
\item {\bf Conjugacy Decision Problem:}
Given $u,w \in B_n$, determine whether they are conjugate,
i.e., there exists $v \in B_n$ such that
$$w=v^{-1}uv$$

\medskip

\item {\bf Conjugacy Search Problem:}
Given conjugate elements\break $u,w \in B_n$, find $v \in B_n$ such that
$$w=v^{-1}uv$$

\medskip

\item {\bf Multiple Simultaneous Conjugacy Search Problem:}\break
Given $m$ pairs of conjugate elements $(u_1,w_1), \dots ,(u_m,w_m) \in B_n$ which are all conjugated by the same element. Find $v \in B_n$ such that
$$w_i=v^{-1} u_i v, \quad \forall i \in \{ 1, \dots, m \}$$

\medskip

\item {\bf Decomposition Problem:} $u \not\in G \leq B_n$. Find $x, y \in G$ such that $w = xuy$.
\end{itemize}

\medskip

\subsection{Key-exchange protocols based on the braid group}

In this section, we present two key-exchange protocols which are
based on apparently hard problems in the braid group.
After the transmitter and receiver agree on a shared secret key,
they can use a symmetric cryptosystem for transmitting messages in the
insecure channel.

\subsubsection{Anshel-Anshel-Goldfeld key-exchange protocol}

The following scheme was proposed theoretically by Anshel, Anshel and Goldfeld \cite{AAG},
and implemented in the braid group by Anshel, Anshel, Fisher and Goldfeld \cite{AAFG}.

This scheme assumes that the Conjugacy Search Problem is difficult enough (so this scheme, as well
as the other schemes described below, would keep its interest, even if it turned out
that braid groups are not relevant, since it might be implemented in other groups).

\medskip

Let $G$ be a subgroup of $B_n$:
$$G=\langle g_1, \dots, g_m \rangle, \qquad g_i \in B_n$$
The secret keys of Alice and Bob are words $a\in G$ and $b \in G$ respectively.

\medskip

The key-exchange protocol is as follows:
\begin{protocol}\ \\

{\em Public keys:} $\{ g_1, \dots , g_m \} \subset B_n$.

{\em Private keys:} Alice: $a$; Bob: $b$.

\medskip

{\em Alice:} Sends Bob publicly the conjugates: $a g_1 a^{-1}, \dots , a g_m a^{-1}$.

{\em Bob:}  Sends Alice publicly the conjugates: $b g_1 b^{-1}, \dots ,  b g_m b^{-1}$.

\medskip

{\em Shared secret key:} $K = aba^{-1}b^{-1}$

\end{protocol}

$K$ is indeed a shared key, since if $a=x_1 \cdots x_k$ where $x_i=g_j^{\pm 1}$ for some $j$, then Alice can compute  $ba^{-1}b^{-1} = (b x_k^{-1} b^{-1})\cdots (b x_1^{-1} b^{-1})$ and hence Alice knows $K=a(ba^{-1}b^{-1})$.
Similarly, Bob can compute  $aba^{-1}$, and hence he knows $K =(aba^{-1})b^{-1}$ .

The security is based on the difficulty of a variant to the Conjugacy Search Problem
in $B_n$, namely the {\it Multiple Conjugacy Search Problem}, in which one tries to find
a conjugating braid starting not from one single pair of conjugate braids $(g, aga^{-1})$,
but from a finite family of such pairs $(g_1, a g_1 a^{-1}), \dots , (g_m, a g_m a^{-1})$ obtained using the same conjugating braid. It should be noted that the Multiple Conjugacy Search Problem may be easier than the original Conjugacy Search Problem.

In \cite{AAFG}, it is suggested to work in $B_{80}$ with $m = 20$ and short initial braids $g_i$ of length $5$ or $10$ Artin generators.

\begin{remark}
We simplified a bit the protocol given by Anshel-Anshel-Goldfeld, but the principle remains the same. Moreover, in their protocol, they used not the braids themselves, but their images   under the colored Burau representation of the braid group defined by Morton \cite{Mor} (see Section \ref{Burau} below).
\end{remark}

\subsubsection{Diffie-Hellman-type key-exchange protocol}\label{DH_braid}

Following the commutative idea for achieving a shared secret key of
Diffie-Hellman, Ko et al. \cite{KLCHKP} propose a key-exchange
protocol based on the braid group and some commutative property of
some of its elements. Although braid groups are not commutative, we
can find large subgroups such that each element of the first
subgroup commutes with each element of the second. Indeed, braids
involving disjoint sets of strands commute. Similar approach appears also
in the Algebraic Eraser Scheme (see \cite{AAGL} and Section \ref{Eraser} here).

Note that this scheme was proposed independently in \cite{SCY} in the context of a general, unspecified noncommutative semigroup with difficult conjugacy problem, but the braid groups were not mentioned there explicitly.

\medskip

Denote by $LB_n$ (resp. $UB_n$) the subgroup of $B_n$ generated
by $\sigma_1, \dots , \sigma_{m-1}$ (resp. $\sigma_{m+1}, \dots , \sigma_{n-1}$) with $m = \lfloor \frac{n}{2} \rfloor$.
Then, every braid in $LB_n$ commutes with every braid in $UB_n$.

Here is Ko et al. key-exchange protocol:

\begin{protocol}\

{\em Public key:} one braid $p$ in $B_n$.

{\em Private keys:} Alice: $s \in LB_n$; Bob: $r \in UB_n$.

\medskip

{\em Alice:} Sends Bob $p' = sps^{-1}$.

{\em Bob:} Sends Alice $p''= rpr^{-1}$

\medskip

{\em Shared secret key:} $K = srpr^{-1}s^{-1}$
\end{protocol}

$K$ is a shared  key since Alice can compute  $K = sp''s^{-1}$ and
Bob can compute  $K = rp'r{-1}$, and both are equal to $K$ since $s$
and $r$ commute.

The security is based on the difficulty of the Conjugacy Search Problem
in $B_n$, or, more exactly, on the difficulty of the following variant, which can be called
the Diffie-Hellman-like Conjugacy Problem:

\begin{problem}
Given a braid $p$ in $B_n$, and the braids
$p' = sps^{-1}$ and $p'' = rpr^{-1}$, where $s \in LB_n$ and $r\in UB_n$, find the
braid $rp'r^{-1}$, which is also $sp''s^{-1}$.
\end{problem}

The suggested parameters are $n=80$, i.e.\ to work in $B_{80}$, with
braids specified using (normal) sequences of length $12$, i.e.,
sequences of $12$ permutation braids (see \cite{CCHKL}).

\subsection{Encryption and decryption}

The following scheme is proposed by Ko et al. \cite{KLCHKP}. We continue with the same notation of Ko et al. Assume that $h$ is a public collision-free
one-way hash function of $B_n$ to $\{0, 1\}^{\mathbb{N}}$, i.e., a computable function such that
the probability of having $h(b_2) = h(b_1)$ for $b_2 \neq b_1$ is negligible (collision-free), and
retrieving $b$ from $h(b)$ is infeasible (one-way) (for some examples see Dehornoy \cite[Section 4.4]{Deh04} and Myasnikov \cite{M08}).

We start with $p \in B_n$ and $s \in LB_n$. Alice's public key is the pair $(p, p')$ with
$p' = sps^{-1}$, where $s$ is Alice's private key. For sending the message $m_B$, which we
assume lies in $\{0, 1\}^{\mathbb{N}}$, Bob chooses a random braid $r$ in $UB_n$ and he sends the encrypted text
$m''_B = m_B \oplus h(rp'r^{-1})$ (using $\oplus$ for the Boolean operation "exclusive-or", i.e.\ the
sum in $\mathbb{Z}/2\mathbb{Z}$), together with the additional datum $p'' = rpr^{-1}$.
Now, Alice computes $m_A = m'' \oplus h(sp''s^{-1})$, and we have $m_A = m_B$, which means that Alice retrieves Bob's original message.

Indeed, because the braids $r$ and $s$ commute, we have (as before):
$$sp''s^{-1} = srpr^{-1}s^{-1} = rsps^{-1}r^{-1} = rp'r^{-1},$$
and, therefore, $m_A = m_B \oplus h(rp'r^{-1}) \oplus h(rp'r^{-1}) = m_B$.

The security
is based on the difficulty of the Diffie-Hellmann-like Conjugacy Problem in $B_n$.
The recommended parameters are as in Ko et al's exchange-key protocol (see Section \ref{DH_braid}).

\subsection{Authentication schemes}

Three authentication schemes were introduced by Sibert, Dehornoy and Girault \cite{SDG02},
which are based on the Conjugacy Search problem and Root Extraction Problem.
Concerning the cryptanalysis of the Root Extraction Problem, see \cite{GHS}.

\medskip

We present here their first scheme. This scheme is related to
Diffie-Hellman based exchange-key in its idea of verifying that the
secret key computed at the two ends is the same.

Note that any encryption scheme can be transformed into an
authentication scheme, by sending to Alice both an encrypted version
and  a hashed image of the same message $m$, then requesting her to
reply with the deciphered message $m$ (she will do it only if the
hashed image of the deciphered message is the same as the one sent
by Bob).

Their first scheme  is based on the difficulty of Diffie-Hellman-like Conjugacy
Problem. It uses the fact that braids involving disjoint families of
strands commute. The data consist of a public key, which is a pair
of braids, and of Alice's private key, also a braid. We assume that
$n$ is even, and denote by $LB_n$ (resp. $UB_n$) the subgroup of
$B_n$ generated by  $\sigma_1, \dots , \sigma_{\frac{n}{2}-1}$,
i.e., braids where the $\frac{n}{2}$ lower strands only are braided
(resp. in the subgroup generated by $\sigma_{\frac{n}{2}+1}, \dots ,
\sigma_{n-1}$). The point is that every element in $LB_n$ commutes
with every element in $UB_n$, and alternative subgroups with this
property could be used instead. We assume that $H$ is a fixed
collision-free hash function from braids to sequences of 0's and 1's
or, possibly, to braids.

\begin{itemize}
\item Phase 1. Key generation:

\begin{enumerate}
\item Choose a public braid $b$ in $B_n$ such that the Diffie-Hellman-like Conjugacy
Problem for $b$ is hard enough;
\item Alice chooses a secret braid $s$ in $LB_n$, her private key; she publishes $b' = sbs^{-1}$;
the pair $(b, b')$ is her public key.
\end{enumerate}

\medskip

\item Phase 2. Authentication phase:

\begin{enumerate}
\item Bob chooses a braid $r$ in $UB_n$, and sends the challenge $x = rbr^{-1}$ to Alice;
\item Alice sends the response $y = H(sxs^{-1})$ to Bob, and Bob checks $y = H(rb'r^{-1})$.
\end{enumerate}

\end{itemize}

For active attacks, the security is ensured by the hash function H: if H is one-way, these
attacks are ineffective.

\medskip

Two more authentication schemes were suggested by Lal and Chaturvedi \cite{LC05}.
Their cryptanalysis are discussed in \cite{GHS,Tsa05}.


\section{Attacks on the conjugacy search problem using Summit Sets}\label{sec_Summit_Sets}

In this section, we explain the algorithms for solving the Conjugacy
Decision Problem and the Conjugacy Search Problem (CDP/CSP) in braid groups which are based on Summit sets.
These algorithms are given in \cite{Gar,ElMo,ECHLPT, FM,Ge,GeGM2}.
We start with the basic idea, and then we continue with its implementations.

We follow here the excellent presentation of Birman, Gebhardt and
Gonz\'alez-Meneses \cite{BGGM1}. For more details, see their paper.

\subsection{The basic idea}
Given an element $x \in B_n$, the algorithm computes a finite subset
$I_x$ of the conjugacy class of $x$ which has the following
properties:
\begin{enumerate}
\item For every $x \in B_n$, the set $I_x$ is finite, non-empty and only
    depends on the conjugacy class of $x$. It means that two elements $x,y \in B_n$
    are conjugate if and only if $I_x = I_y$.

\item For each $x\in B_n$, one can compute efficiently a representative $\tilde x \in I_x$ and an element $a \in B_n$ such that $a^{-1} x a = \tilde x$.

\item There is a finite algorithm which can construct the whole set $I_x$ from any
    representative $\tilde x \in I_x$.
\end{enumerate}

Now, for solving the CDP/CSP for given $x, y \in B_n$ we have to
perform the following steps.

\begin{enumerate}
\item[(a)] Find representatives $\tilde{x} \in I_x$ and $\tilde{y} \in I_y$.

\item[(b)] Using the algorithm from property (3), compute further
elements of $I_x$ (while keeping track of the conjugating elements),
until either:
\begin{enumerate}
\item[(i)] $\tilde y$ is found as an element of $I_x$, proving $x$ and $y$
to be conjugate and providing a conjugating element, or

\item[(ii)] the entire set $I_x$ has been constructed without encountering
    $\tilde y$, proving that $x$ and $y$ are not conjugate.
\end{enumerate}
\end{enumerate}

We now survey the different algorithms based on this approach.

\medskip

In Garside's original algorithm \cite{Gar}, the set $I_x$ is the
{\it Summit Set} of $x$, denoted ${\rm SS}(x)$, which is the set of
conjugates of $x$ having maximal infimum.

\begin{remark}
All the algorithms presented below for the different types of Summit Sets work also for Garside groups (defined by Dehornoy and Paris  \cite{DePa}), which are a generalization of the braid groups. In our survey, for
simplification, we present them in the language of braid groups. For
more details on the Garside groups and the generalized algorithms,
see \cite{BGGM1}.
\end{remark}

\subsection{The Super Summit Sets}
The Summit Set are improved by El-Rifai and Morton \cite{ElMo}, who
consider $I_x = {\rm SSS}(x)$, the {\it Super Summit Set} of $x$,
consisting of the conjugates of $x$ having minimal canonical length
$\ell(x)$. They also show that ${\rm SSS}(x)$ is the set of
conjugates of $x$ having maximal infimum and minimal supremum, at
the same time. El-Rifai and Morton \cite{ElMo} show that ${\rm SSS}(x)$ is finite.
In general, ${\rm SSS}(x)$ is much smaller than ${\rm SS}(x)$.
For example, take the element $x = \Delta_4 \sigma_1 \sigma_1 \in B_4$,
$\SSS(x) = \{ \Delta_4 \cdot \sigma_1 \sigma_3 \}$ while
$${\rm SS}(x) = \{ \Delta_4 \cdot \sigma_1 \sigma_3, \Delta_4 \cdot \sigma_1 \cdot \sigma_1,
\Delta_4 \cdot \sigma_3 \cdot \sigma_3 \}$$ (the factors
in each left normal form are separated by a dot) \cite[page 8]{BGGM1}.

Starting by a given element $x$, one can find an element $\tilde x
\in  {\rm SSS}(x)$ by a sequence of special conjugations, called
{\it cyclings} and {\it decyclings}:

\begin{defn}
Let $x =\Delta^p x_1 \cdots x_r \in B_n$ be given in Garside's normal
form and assume $r>0$.

The {\em cycling} of $x$, denoted by ${\bf c}(x)$ is:
$${\bf c}(x) = \Delta ^p x_2 \cdots x_r \tau^{-p} (x_1),$$
where $\tau$ is the involution which maps $\sigma_i$ to
$\sigma_{n-i}$, for all $1 \leq i \leq n$.

The {\em decycling} of $x$, denoted by ${\bf d}(x)$ is:
$${\bf d}(x) = x_r \Delta ^p x_1 x_2 \cdots x_{r-1}
= \Delta ^p \tau^{p} (x_r) x_1 x_2 \cdots x_{r-1}.$$

If $r=0$, we have ${\bf c}(x) = {\bf d}(x) = x$.
\end{defn}

Note that ${\bf c}(x) = (\tau^{-p} (x_1))^{-1} x (\tau^{-p} (x_1))$
and ${\bf d}(x) = x_r^{-1} x x_r$. This means that for an element of
positive canonical length, the cycling of $x$ is computed by moving
the first permutation braid of $x$ to the end, while the decycling of
$x$ is computed by moving the last permutation braid of $x$ to the
front. Moreover, for every $x \in B_n$, ${\rm inf}(x) \leq {\rm
inf}({\bf c}(x))$ and ${\rm sup}(x) \geq {\rm sup}({\bf d}(x))$.

Note that the above decompositions of ${\bf c}(x)$ and ${\bf
d}(x)$ are not, in general, Garside's normal forms. Hence, if one
wants to perform iterated cyclings or decyclings, one needs to
compute the left normal form of the resulting element at each
iteration.

Given $x$, one can use cyclings and decyclings to find an element in
${\rm SSS}(x)$ in the following way: Suppose that we have an element
$x \in B_n$ such that ${\rm inf}(x)$ is not equal to the maximal
infimum in the conjugacy class of $x$. Then, we can increase the
infimum by repeated cycling (due to \cite{BKL01,ElMo}):
there exists a positive integer $k_1$ such that ${\rm inf}({\bf c}^{k_1}
(x)) > {\rm inf}(x)$. Therefore, by repeated cycling, we can
conjugate $x$ to another element $\hat{x}$ of maximal infimum. Once
$\hat{x}$ is obtained, if the supremum is not minimal in the
conjugacy class, we can decrease its supremum by repeated decycling.
Again, due to  \cite{BKL01,ElMo}, there exists an integer
$k_2$ such that ${\rm sup}({\bf d}^{k_2}(\hat{x})) < {\rm sup}(\hat{x})$.
Hence, using repeated cycling and decycling a finite number of
times, one obtains an element in $\SSS(x)$.

If we denote by $m$ the length of $\Delta$ in Artin generators and $r$ is
the canonical length of $x$, then we have (see \cite{BKL01,ElMo}):
\begin{prop}
A sequence of at most $rm$ cyclings and decyclings applied to $x$
produces a representative $\tilde{x} \in \SSS(x)$.
\end{prop}

Now, we have to explore all the set $\SSS(x)$. We have the following
result (see \cite{ElMo}):
\begin{prop}
Let $x \in B_n$ and $V \subset \SSS(x)$ be non-empty. If $V \neq
\SSS(x)$, then there exist $y \in V$  and a permutation braid $s$
such that $s^{-1}y s  \in \SSS(x) \setminus V$.
\end{prop}

Since $\SSS(x)$ is a finite set, the above proposition allows us to
compute the whole $\SSS(x)$. More precisely, if one knows a subset
$V \subset \SSS(x)$ (we start with: $V = \{ \tilde{x} \}$), one
conjugates each element in $V$ by all permutation braids ($n!$
elements). If one encounters a new element $z$ with the same
canonical length as $\tilde{x}$ (which is a new element in
$\SSS(x)$), then add $z$ to $V$ and start again. If no new element
is found, this means that $V = \SSS(x)$, and we are done.

One important remark is that this algorithm not only computes the
set $\SSS(x)$, but it also provides conjugating elements joining the
elements in $\SSS(x)$.

Now the checking if $x$ and $y$ are conjugate,  is done as
follows: Compute representatives $\tilde{x} \in \SSS(x)$ and
$\tilde{y} \in \SSS(y)$. If ${\rm inf}(\tilde{x}) \neq {\rm inf}(\tilde{y})$ or
${\rm sup}(\tilde{x}) \neq {\rm sup}(\tilde{y})$, then $x$ and $y$ are not conjugate.
Otherwise, start computing $\SSS(x)$ as described above. The
elements $x$ and $y$ are conjugate if and only if $\tilde{y} \in
\SSS(x)$. Note that if $x$ and $y$ are conjugate, an element
conjugating $x$ to $y$ can be found by keeping track of the
conjugations during the computations of $\tilde{x}$, $\tilde{y}$ and
$\SSS(x)$. Hence, it solves the Conjugacy Decision Problem and the
Conjugacy Search Problem simultaneously.

From the algorithm, we see that the computational cost of computing
$\SSS(x)$ depends mainly in two ingredients: the size of $\SSS(x)$
and the number of permutation braids. In $B_n$, all known upper
bounds for the size of $\SSS(x)$ are exponential in $n$, although it
is conjectured that for fixed $n$, a polynomial bound in the
canonical length of $x$ exists \cite{ECHLPT}.

\medskip

Franco and Gonz\'alez-Meneses \cite{FM} reduce the size of the set we have to conjugate with, by the
following observation:

\begin{prop}
Let $x \in B_n$ and $y \in  \SSS(x)$. For every positive braid $u$
there is a unique $\preceq$-minimal element $c_y (u)$ satisfying $u
\preceq c_y (u)$ and $(c_y(u))^{-1} y (c_y (u)) \in \SSS(x)$.
\end{prop}

\begin{defn}
Given $x \in B_n$ and $y \in \SSS(x)$, we say that a permutation
braid $s \neq 1$ is {\em minimal} for $y$ with respect to
$\SSS(x)$ if $ s^{-1}y s \in \SSS(x)$, and no proper prefix of $s$
satisfies this property.
\end{defn}

It is easy to see that the number of minimal permutation braids for
$y$ is bounded by the number of Artin's generators.

Now, we have:

\begin{prop}
Let $x \in B_n$ and $V \subseteq \SSS(x)$ be non-empty. If $V \neq
\SSS(x)$, then there exist $y \in V$ and a generator $\sigma_i$ such
that $c_y (\sigma_i)$ is a minimal permutation braid for $y$, and
$(c_y (\sigma_i))^{-1} y (c_y (\sigma_i)) \in \SSS(X)\setminus V$.
\end{prop}

Using these proposition, the $\SSS(x)$ can be computed as in
\cite{ElMo}, but instead of conjugating each element $y \in \SSS(x)$
by all permutation braids ($n!$ elements), it suffices to conjugate $y$ by the minimal permutation braids $c_y(\sigma_i)$ ($1 \leq i \leq n-1$, $n-1$ elements).

\begin{figure}[!ht]
\epsfysize=6cm \centerline{\epsfbox{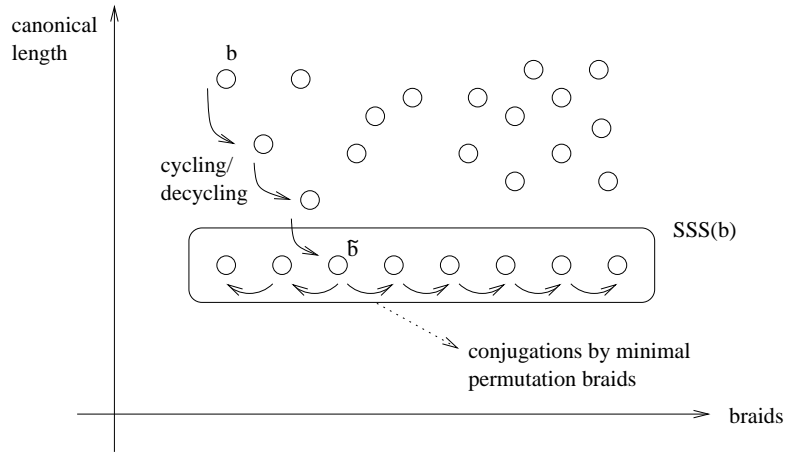}}
\caption{Solving the conjugacy problem: going to the SSS and then
exploring it (the points represent the conjugates of $b$)}\label{SSS_complexity_graph}
\end{figure}

Figure \ref{SSS_complexity_graph} (taken from \cite{Deh04}) summarizes the solution of the conjugacy problem using the Super Summit Set for an element $b$.

Note that the algorithm computes a directed graph whose vertices are
the elements in $\SSS(x)$, and whose arrows are defined as follows:
for any two elements $y,z \in \SSS(x)$, there is an arrow labeled by
the minimal permutation braid $p_i$ starting at $y$ and ending at $z$ if
$p_i^{-1} y p_i = z$.

\begin{figure}[!htf]
\epsfysize=3cm \centerline{\epsfbox{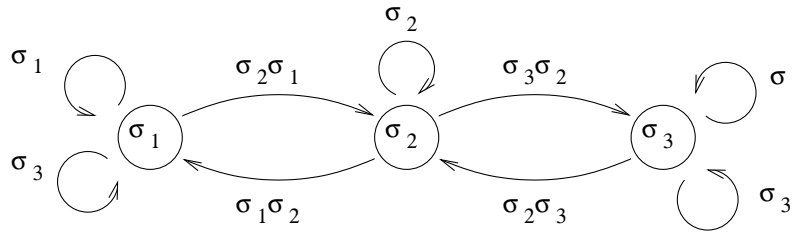}}
\caption{The graph of $\SSS(\sigma_1)$ in $B_4$}\label{SSS_graph}
\end{figure}

An example for such a graph can be seen in Figure \ref{SSS_graph}, for the set $\SSS(\sigma_1)$ in $B_4$ (taken from \cite[pp. 10--11]{BGGM1}).
Note that there are exactly $3$ arrows starting at every vertex (the number of Artin generators of $B_4$). In general, the number of arrows starting at a
given vertex can be smaller or equal, but never larger than the number of generators.

Hence, the size of the set of permutation braids is no longer a
problem for the complexity of the algorithm (since we can use the
minimal permutation braids instead), but there is still a big problem to
handle: The size of $\SSS(x)$ is, in general, very big. The next
improvement tries to deal with this.

\subsection{The Ultra Summit Sets}

Gebhardt \cite{Ge} defines a small subset of $\SSS(x)$ satisfying
all the good properties described above, so that a similar algorithm
can be used to compute it. The definition of this new subset
appears after observing that the cycling function maps $\SSS(x)$ to
itself. As $\SSS(x)$ is finite, iterated cycling of any
representative of $\SSS(x)$ must eventually become periodic. Hence
it is natural to define the following:

\begin{defn}
Given $x \in B_n$, the {\em Ultra Summit Set} of $x$,
$\USS(x)$, is the set of elements $y \in \SSS(x)$ such that ${\bf
c}^m(y ) = y$ for some $m> 0$.
\end{defn}

\begin{figure}[!htf]
\epsfysize=4cm \centerline{\epsfbox{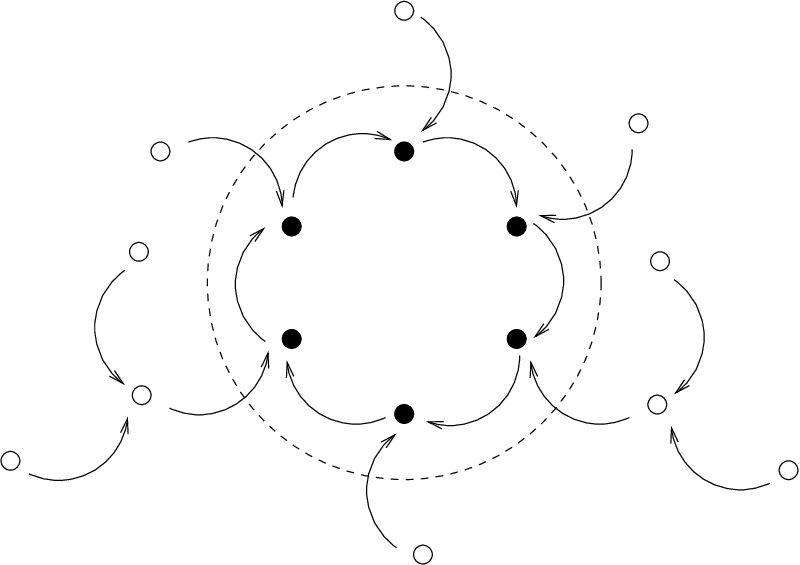}}
\caption{Action of cycling inside the Super Summit Set; the elements of the Ultra Summit Set are in black and perform some orbits under cycling
(taken from \cite[Figure 4]{Deh04})}\label{USS_cycle_graph}
\end{figure}

Hence, the Ultra Summit Set $\USS(x)$ consists of a finite set of
disjoint orbits, closed under cycling (see some schematic example in Figure \ref{USS_cycle_graph}).

\begin{example}\label{exam_rigid}\cite{BGGM1}
One has
$$\USS(\sigma_1) = \SSS(\sigma_1) ={\rm SS}(\sigma_1) =
\{\sigma_1, \dots, \sigma_{n-1}\},$$ and each element corresponds to
an orbit under cycling, since ${\bf c}(\sigma_i) = \sigma_i$ for $i = 1,
\dots , n-1$.

A more interesting example is given by the element
$$x = \sigma_1\sigma_3\sigma_2\sigma_1 \cdot \sigma_1 \sigma_2 \cdot \sigma_2 \sigma_1 \sigma_3 \in B_4.$$
In this example, $\USS(x)$ has $6$ elements, while $\SSS(x)$ has
$22$ elements. More precisely, $\USS(x)$ consists
of 2 closed orbits under cycling: $\USS(x) = O_1 \cup O_2$, each one
containing $3$ elements:
$$O_1 = \left\{\begin{array}{c}
\sigma_1\sigma_3\sigma_2\sigma_1 \cdot \sigma_1\sigma_2 \cdot \sigma_2\sigma_1\sigma_3,\\
\sigma_1\sigma_2 \cdot \sigma_2\sigma_1\sigma_3 \cdot \sigma_1\sigma_3\sigma_2\sigma_1,\\
\sigma_2\sigma_1\sigma_3 \cdot \sigma_1\sigma_3\sigma_2\sigma_1 \cdot \sigma_1\sigma_2
\end{array}\right\},$$
$$O_2 = \left\{ \begin{array}{c}
\sigma_3\sigma_1\sigma_2\sigma_3 \cdot \sigma_3\sigma_2 \cdot \sigma_2\sigma_3\sigma_1,\\
\sigma_3\sigma_2 \cdot \sigma_2\sigma_3\sigma_1 \cdot \sigma_3\sigma_1\sigma_2\sigma_3,\\
\sigma_2\sigma_3\sigma_1 \cdot \sigma_3\sigma_1\sigma_2\sigma_3 \cdot \sigma_3\sigma_2
\end{array}\right\}.$$
Notice that $O_2 = \tau(O_1)$.

Note also that the cycling of every element in $\USS(x)$ gives
another element which is already in left normal form, hence iterated
cyclings corresponds to cyclic permutations of the factors in the
left normal form. Elements which satisfies this property are called
{\em rigid} (see \cite{BGGM1}).
\end{example}

\begin{remark}
The size of the Ultra Summit Set of a {\it generic} braid of
canonical length $\ell$ is either $\ell$ or $2\ell$ \cite{Ge}. This
means that, in the generic case, Ultra Summit Sets consist of one or
two orbits (depending on whether $\tau(O_1) = O_1$ or not),
containing rigid braids. But, there are exceptions: for example, the
following braid in $B_{12}$:
\begin{eqnarray*}
E & = & (\sigma_2\sigma_1\sigma_7\sigma_6\sigma_5\sigma_4\sigma_3\sigma_8\sigma_7\sigma_{11}\sigma_{10}) \cdot
(\sigma_1\sigma_2\sigma_3\sigma_2\sigma_1\sigma_4\sigma_3\sigma_{10}) \cdot \\
& & \cdot (\sigma_1\sigma_3\sigma_4\sigma_{10}) \cdot
(\sigma_1\sigma_{10}) \cdot
(\sigma_1\sigma_{10}\sigma_9\sigma_8\sigma_7\sigma_{11}) \cdot
(\sigma_1\sigma_2\sigma_7\sigma_{11})
\end{eqnarray*}
has an Ultra Summit Set of size $264$, instead of the expected size
$12$ (see \cite[Example 5.1]{BGGM2}).

In the case of braid groups, the size
and structure of the Ultra Summit Sets happen to depend very much on the
geometrical properties of the braid, more precisely, on its
Nielsen-Thurston type: periodic, reducible or Pseudo-Anosov (see \cite{BGGM1,BGGM2}).
\end{remark}

The algorithm given in \cite{Ge} to solve the CDP/CSP in braid
groups (using Ultra Summit Sets) is analogous to the previous ones,
but this time one needs to compute $\USS(x)$ instead of $\SSS(x)$.
In order to do this, we first have to obtain an element $\hat{x} \in
\USS(x)$.  We do this as follows: take an element $\tilde{x} \in
\SSS(x)$. Now, start cycling it. Due to the facts that cycling an
element in $\SSS(x)$ will result in another element in $\SSS(x)$ and
that the Super Summit Set of $x$ is finite, we will have two integers
$m_1,m_2$ ($m_1 < m_2$), which satisfy:
$${\bf c}^{m_1}(\tilde{x})={\bf c}^{m_2}(\tilde{x})$$
When having this, the element $\hat{x}={\bf c}^{m_1}(\tilde{x})$ is
in $\USS(x)$, since: $${\bf c}^{m_2-m_1}(\hat{x})=\hat{x}.$$

After finding a representative $\hat{x} \in \USS(x)$, we have to
explore all the set $\USS(x)$. This we do using the following
results of Gebhardt \cite{Ge} (which are similar to the case of the
Super Summit Set):

\begin{prop}
Let $x \in B_n$ and $y \in  \USS(x)$. For every positive braid $u$
there is a unique $\preceq$-minimal element $c_y (u)$ satisfying $u
\preceq c_y (u)$ and $(c_y(u))^{-1} y (c_y (u)) \in \USS(x)$.
\end{prop}

\begin{defn}
Given $x \in B_n$ and $y \in \USS(x)$, we say that a permutation
braid $s \neq 1$ is a {\em minimal} for $y$ with respect to
$\USS(x)$ if $ s^{-1}y s \in \USS(x)$, and no proper prefix of $s$
satisfies this property.
\end{defn}

It is easy to see that the number of minimal permutation braids for
$y$ is bounded by the number of Artin's generators.

Now, we have:

\begin{prop}
Let $x \in B_n$ and $V \subseteq \USS(x)$ be non-empty. If $V \neq
\USS(x)$, then there exist $y \in V$ and a generator $\sigma_i$ such
that $c_y (\sigma_i)$ is a minimal permutation braid for $y$, and
$(c_y (\sigma_i))^{-1} y (c_y (\sigma_i)) \in \USS(X)\setminus V$.
\end{prop}

In \cite{Ge}, it is shown how to compute the minimal permutation
braids (they  are called there {\it minimal simple elements} in the
Garside group's language) corresponding to a given $y \in \USS(x)$
(a further discussion on the minimal simple elements with some examples
can be found in \cite{BGGM2}).
Hence, one can compute the whole $\USS(x)$ starting by a single
element $\hat{x} \in \USS(x)$, and then we are done.

\medskip

For a better characterization of the minimal permutation braids, let us introduce some notions related to a braid given in a left normal form (see \cite{BGGM2}):
\begin{defn}\label{initial-final}
Given $x \in B_n$ whose left normal form is\break $x = \Delta^p x_1 \cdots x_r \ (r > 0)$, we define the {\em initial factor of $x$} as\break $\iota(x) = \tau^{-p}(x_1)$,
and the {\em final factor of $x$} as $\varphi(x) = x_r$. If $r = 0$ we define
$\iota(\Delta^p) = 1$ and $\varphi(\Delta^p) = \Delta$.
\end{defn}

\begin{defn}
Let $u, v$ be permutation braids such that $uv = \Delta$. The {\em right complement of $u$}, $\partial(u)$, is defined by $\partial(u)=u^{-1}\Delta = v$.
\end{defn}

Note that a cycling of $x$ is actually a conjugation of $x$ by the initial factor $\iota(x)$: ${\bf c}(x) =\iota(x)^{-1} x \iota(x)$, and a decycling of $x$ is actually a conjugation of $x$ by the inverse of final factor $\varphi(x)^{-1}$:
${\bf d}(x) =\varphi(x) x \varphi(x)^{-1}$.

\medskip

The notions of Definition \ref{initial-final} are closely related (see \cite{BGGM1}):
\begin{lemma}
For every $x \in B_n$, one has $\iota(x^{-1}) = \partial(\varphi(x))$ and
$\varphi(x^{-1}) =\partial^{-1}(\iota(x))$.
\end{lemma}

The following proposition from \cite{BGGM2} characterizes the minimal permutation braids for $x$ as prefixes of $x$  or of $x^{-1}$:

\begin{prop}
Let $x \in \USS(x)$ with $\ell(x) > 0$ and let $c_x(\sigma_i)$ be a minimal permutation braid for $x$.
Then $c_x(\sigma_i)$ is a prefix of either $\iota(x)$ or $\iota(x^{-1})$, or both.
\end{prop}

As in the case of the Super Summit Set, the algorithm of Gebhardt
\cite{Ge} not only computes $\USS(x)$, but also a graph $\Gamma_x$,
which determines the conjugating elements. This graph is defined as
follows.

\begin{defn}
Given $x \in B_n$, the directed graph $\Gamma_x$ is defined by the
following data:
\begin{enumerate}
\item The set of vertices is $\USS(x)$.

\item For every $y \in \USS(x)$ and every minimal permutation braid $s$ for
$y$ with respect to $\USS(x)$, there is an arrow labeled by $s$
going from $y$ to $s^{-1}y s$.
\end{enumerate}
\end{defn}

\begin{example}
Let us give some example for the graph $\Gamma_x$. We follow \cite[Example 2.10]{BGGM2}.

Let $x = \sigma_1\sigma_2\sigma_3\sigma_2 \cdot \sigma_2 \sigma_1\sigma_3 \cdot \sigma_1\sigma_3\in B_4$.
This braid A is Pseudo-Anosov and rigid. A computation shows that $\USS(x)$ has
exactly two cycling orbits, with $3$ elements each, namely:
$$x_1 = \left\{ \begin{array}{c}
x_{1,1} = \sigma_1\sigma_2\sigma_3\sigma_2 \cdot \sigma_2\sigma_1\sigma_3 \cdot \sigma_1\sigma_3,\\
x_{1,2} = \sigma_2\sigma_1\sigma_3 \cdot \sigma_1\sigma_3 \cdot \sigma_1\sigma_2\sigma_3\sigma_2,\\
x_{1,3} = \sigma_1\sigma_3 \cdot \sigma_1\sigma_2\sigma_3\sigma_2 \cdot \sigma_2\sigma_1\sigma_3 \end{array}\right\},
$$

$$x_2 = \left\{ \begin{array}{c}
x_{2,1} = \sigma_1\sigma_3\sigma_2\sigma_1 \cdot \sigma_2\sigma_1\sigma_3 \cdot \sigma_1\sigma_3,\\
x_{2,2} = \sigma_2\sigma_1\sigma_3 \cdot \sigma_1\sigma_3 \cdot \sigma_1\sigma_3\sigma_2\sigma_1,\\
x_{2,3} = \sigma_1\sigma_3 \cdot \sigma_1\sigma_3\sigma_2\sigma_1 \cdot \sigma_2\sigma_1\sigma_3 \end{array}\right\}.
$$

The graph $\Gamma_x$ of $\,\USS(x)$ is illustrated in Figure \ref{USS_graph}.
The solid arrows are conjugations by minimal permutation braids which are prefixes of the initial factors, while the dashed arrows are conjugations by minimal permutation braids which  are prefixes of the final factors. Note that the definitions imply that the cycles $x_1$ and $x_2$ of $\,\USS(x)$ are connected by solid arrows.

\begin{figure}[!htf]
\epsfysize=8cm \centerline{\epsfbox{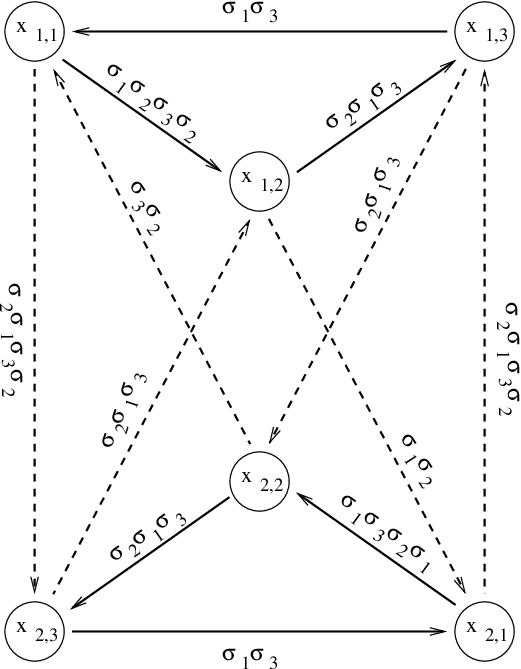}}
\caption{The graph of $\USS(\sigma_1\sigma_2\sigma_3\sigma_2\sigma_2 \sigma_1\sigma_3\sigma_1\sigma_3) \subset B_4$}\label{USS_graph}
\end{figure}

\end{example}

Concerning the complexity of this algorithm for solving the Conjugacy Search
Problem, the number $m_2$ of times one needs to apply cycling for
finding an element in $\USS(x)$ is not known in general.
Nevertheless, in practice, the algorithm based on the Ultra Summit
Set is substantially better for braid groups (see \cite{BGGM1}). For
more information on the Ultra Summit Set and its structure, see
\cite{BGGM2}.

\begin{remark}
One might think that for a given element $x\in B_n$, it is possible
that its Ultra Summit Set with respect to the Garside normal form
will be different from its Ultra Summit Set with respect to the {\it
right normal form} (see Section \ref{GNF_Sec}). If this happens, it is possible that even
though one of the Ultra Summit Sets is large, the other will be
small.

Gebhardt and Gonz\'alez-Meneses \cite{GeGM} show that at least for {\em rigid
braids}, the size of the above two Ultra Summit Sets is equal, and
their associated graphs are isomorphic (a braid $w$ is called {\em
rigid}, if the cycling of $w$,\ ${\bf c}(w)$, is already given in
Garside normal form, with no need for changing the permutation
braids; see also \cite[Section 3]{BGGM1} and Example \ref{exam_rigid} here).
They conjecture that this is the situation for any braid.
\end{remark}

\subsection{Some variants of the Ultra Summit Sets}

In this section, we sketch some variants of the Super Summit Sets and the Ultra Summit Sets suggested by several authors.

\subsubsection{Reduced Super Summit Sets}
Lee, in his thesis \cite{Lee} (2000), suggests a variant of the Super Summit Set, which is actually a subset of the Ultra Summit Set which was defined later (2005) by Gebhardt:

\begin{defn}
The {\em Reduced Super Summit Set of $x$}, denote by ${\rm RSSS}(x)$, is:
$${\rm RSSS}(x) = \{ y \in C(x)| {\bf c}^m(y) = y\ {\rm and}\ {\bf d}^n(y) = y\
{\rm for\ some}\ m, n \geq 1 \}.$$
where $C(x)$ is the conjugacy class of $x$
\end{defn}

Lee's motivation to look on ${\rm RSSS}(x)$ comes from the facts that it is still easy to find algorithmically an element in ${\rm RSSS}(x)$ for a given $x$, this set is invariant under cyclings and decyclings, and this set is usually smaller than $\SSS(x)$.

Indeed, it is easy to see (by \cite{ElMo} and \cite{GeGM2}) that:
$${\rm RSSS}(x) \subseteq \USS (x) \subseteq \SSS(x)$$

Lee indicates that there is no known algorithm to generate ${\rm RSSS}(x)$ without generating $\SSS(x)$ before. Despite this, he has succeeded to compute ${\rm RSSS}(x)$ in polynomial time for the case of rigid braids in $B_4$.

\subsubsection{A general cycling operation and its induced set}

Zheng \cite{Zhe} suggests to generalize the idea of cyclings. He defines:
\begin{defn}
The {\em cycling operation of order $q$ on $x$} is the conjugation
${\bf c}_q(x) = s^{-1}xs$, where $s$ is the maximal common prefix of $x$ and $\Delta^q$.
(this will be denoted in the next section as: $s=x \wedge \Delta^q$).

The corresponding set is:
$$G_q = \{x \in B_n \ |\ {\bf c}_q^N (x) = x\ {\rm for\ some\ } N > 0\}.$$
\end{defn}

The new cycling operations are indeed natural generalizations of the cycling
and decycling operation:
$${\bf c}(x) = \tau^{-{\rm inf}(x)} \left({\bf c}_{{\rm inf}(x)+1}(x)\right), \qquad
{\bf d}(x) = {\bf c}_{{\rm sup}(x)-1}(x).$$

 Recall that $C(x)$ is the conjugacy class of $x$. For getting the Super Summit Sets and the Ultra Summit Sets in the language of $G_q$, we define:
$${\rm inf}_s (x) = {\rm max} \{ {\rm inf} (y)\ |\ y \in C(x) \},\qquad
  {\rm sup}_s (x) = {\rm min} \{ {\rm sup} (y)\ |\ y \in C(x) \}.$$
Hence, we get that:
$$\SSS(x) = C(x) \cap \left( {\bigcap_{q \in \{ {\rm inf}_s (x),{\rm sup}_s (x) \} }}
G_q \right),$$
$$\USS(x) = C(x) \cap \left( {\bigcap_{q\in \{ {\rm inf}_s (x),{\rm inf}_s (x)+1,{\rm sup}_s (x) \} }} G_q \right).$$

Zheng \cite{Zhe} defines a new summit set:
$$C^{*}(x) = C(x) \cap  \left( {\bigcap_{q \in \Z}} G_q \right) = C(x) \cap \left( {\bigcap_{{\rm inf}_s (x) \leq q \leq {\rm sup}_s (x)}} G_q \right).$$
It is straight-forward that:
$$C^*(x) \subseteq \USS(x) \subseteq \SSS(x).$$

Given an element $x$, computing an element $\hat{x} \in C^*(x)$ is done by applying iterated general cyclings ${\bf c}_q$ until getting repetitions, for
${\rm inf}(x)< q< {\rm sup}(x)$.
A more complicated algorithm is presented for finding the whole $C^*(x)$ (see \cite[Algorithm 3.8]{Zhe}). Having these ingredients for $C^*(x)$, we can solve the Conjugacy Search Problem based on $C^*(x)$.

Zheng \cite[Section 6]{Zhe} presents some computational results, and he emphasizes that
the new set $C^{*}(x)$ is important especially for the case of reducible braids, where there are cases that $\USS(x)=\SSS(x)$.

\subsubsection{Stable Super Summit Sets and Stable Ultra Summit Sets}

The stable Super Summit Sets and stable Ultra Summit Sets were defined simultaneously by Birman, Gebhardt and Gonz\'alez-Meneses \cite{BGGM1} and Lee and Lee \cite{LeeLee3a}:

\begin{defn}
Given $x \in B_n$,
The {\em stable Super Summit Set of $x$} is defined as:
$${\rm SSSS}(x) = \{ y \in \USS(x)\ |\  y^m \in \USS(x^m), \forall m \in \Z \}.$$
The {\em stable Ultra Summit Set of $x$} is defined as:
$${\rm SU}(x) = \{ y \in \USS(x)\ |\  y^m \in \USS(x^m), \forall m \in \Z \}.$$
\end{defn}

Birman, Gebhardt and Gonz\'alez-Meneses \cite[Proposition 2.23]{BGGM1} and Lee and Lee \cite[Theorem 6.1(i)]{LeeLee3a} have proved that for every $x \in B_n$ the stable sets ${\rm SSSS}(x)$ and ${\rm SU}(x)$ are non-empty.

We give here an example from \cite{LeeLee3a}, which shows that: (i) the stable Super Summit Set is different from both the
Super Summit Set and the Ultra Summit Set; (ii) one cannot obtain an element
of the stable Super Summit Set by applying only cyclings and decyclings.

\begin{example}\cite[page 11]{LeeLee3a}
Consider the positive 4-braid monoid $B_4^+$. Let
$$g_1 = \sigma_1\sigma_2\sigma_3,\quad g_2 = \sigma_3\sigma_2\sigma_1, \quad g_3 = \sigma_1\sigma_3\sigma_2,\quad g_4 = \sigma_2\sigma_1\sigma_3.$$
Note that $g_i$'s are permutation braids and conjugate to each other.

It is easy to see that
$$\SSS(g_1) = \USS(g_1) = \{g_1, g_2, g_3, g_4 \}.$$
Now, we show that the stable Super Summit Set of $g_1$ is different from the Super/Ultra Summit Set of $g_1$. The normal forms of $g_i^2$ are as follows:
$$g_1^2 = (\sigma_1\sigma_2\sigma_3\sigma_1\sigma_2)\sigma_3;\quad
g_2^2 =(\sigma_3\sigma_2\sigma_1\sigma_3\sigma_2)\sigma_1;\quad g_3^2=\Delta; \quad
g_4^2=\Delta.$$
Therefore,
${\rm inf}(g_1^2) = {\rm inf}(g_2^2) = 0$ and ${\rm inf}(g_3^2) = {\rm inf}(g_4^2)=1$.
Hence,
$${\rm SSSS}(g_1) = \{ g_3, g_4\}.$$
Note that ${\bf c}^k(g_i) = {\bf d}^k(g_i) = g_i$ for $i = 1,\dots , 4$ and all $k > 1$. In particular, we cannot obtain
an element of the stable Super Summit Set by applying only cyclings and decyclings to $g_1$ or $g_2$.
\end{example}

A finite-time algorithm for computing the stable Super Summit Sets (i.e. when given $x\in B_n$, first compute an element $\hat{x} \in {\rm SSSS}(x)$ and then compute the whole set ${\rm SSSS}(x)$) is given by Lee and Lee in \cite[Section 6]{LeeLee4}.

Birman, Gebhardt and Gonz\'alez-Meneses \cite[page 27]{BGGM1} remark that their proof for the non-emptiness of the stable Ultra Summit Set (Proposition 2.23 there) actually yields
an algorithm for computing this set.

\medskip

Zheng \cite{Zhe}, as a continuation of his idea of general cyclings, suggests to
generalize also the stable sets. He defines:
\begin{defn}
${\bf c}_{p,q}(x) = s^{-1}xs$, where $s$ is the maximal common prefix of $x^p$ and $\Delta^q$ (i.e., $s=x^p \wedge \Delta^q$).

The corresponding set is:
$$G_{p,q} = \{x \in B_n \ |\ {\bf c}_{p,q}^N (x) = x\ {\rm for\ some\ } N > 0\}.$$
\end{defn}

Note that ${\bf c}_q(x^p) = ({\bf c}_{p,q}(x))^p$, so applying a ${\bf c}_q$ operation on $x^p$ is equivalent
to applying a ${\bf c}_{p,q}$ operation on $x$. In particular, $x^p \in G_q$ if and only if
$x \in G_{p,q}$.

Similarly, one can define:
$$C^{[m,n],*}(x) = C(x) \cap \left( {\bigcap_{m \leq p \leq n, q \in \Z}}
G_{p,q} \right).$$

Zheng claims, that with a suitable modification, the algorithms for computing $C^*(x)$
can be used to compute the set $C^{[m,n],*}(x)$.

An even more generalized set is:
$$C^{*,*}(x) = C(x) \cap \left( {\bigcap_{p,q \in \Z}} G_{p,q} \right),$$
but currently there is no algorithm for computing it, because he does not know how to bound the order $p$. Nevertheless, Zheng \cite[Theorem 7.3]{Zhe} have proved that the set $C^{*,*}(x)$ is nonempty.

The set $C^{*,*}(x)$ is indeed a generalization of the stable sets, since:
$${\rm SSSS}(x)= C(x) \cap \left({\bigcap_{p\geq 1, q \in \{ {\rm inf}_s (x^p),{\rm sup}_s (x^p) \} }} G_{p,q}\right),$$
$${\rm SU}(x)= C(x) \cap \left({\bigcap_{p\geq 1, q \in \{ {\rm inf}_s (x^p), {\rm inf}_s (x^p) +1, {\rm sup}_s (x^p) \}}} G_{p,q}\right).$$

By the non-emptiness result of Zheng, we have an alternative proof that the stable sets are nonempty.

\subsection{Cyclic sliding}\label{subsec_sliding}
The last step up-to-date for seeking a polynomial-time solution to the conjugacy search problem has been done by Gebhardt and Gonz\'alez-Meneses \cite{GeGM2,GeGM3}.

Their idea is introducing a new operation, called {\it cyclic sliding}, and they suggest to replace the usual cycling and decycling operations by this new one,
as it is more natural from both the theoretical and computational points of view.
Then, the Ultra Summit Set $\USS(x)$ of $x$, will be replaced by its analogue for cyclic sliding:  the set of {\em sliding circuits}, ${\rm SC}(x)$. The sets of sliding circuits and their elements naturally satisfy all the good properties that were
already shown for Ultra Summit Sets, and sometimes even better properties: For example,
for elements of canonical length $1$, cycling and decycling are trivial operations, but cyclic sliding is not.

One more advantage of considering the set ${\rm SC}(x)$ is that it yields a simpler algorithm to solve the Conjugacy Decision Problem and the
Conjugacy Search Problem in the braid group. The worst case complexity of the algorithm is not better than the previously known ones \cite{Ge}, but it is conceptually simpler and easier to implement. The details of the implementation and the study of complexity
are presented in \cite{GeGM3}.

\medskip

For any two braids $u,v$, let us denote $u \wedge v$ to be the largest common
prefix of $u$ and $v$ (the notation comes from the corresponding operation on the lattice generated by the partial order $\preceq$ on the elements of $B_n$, see Section \ref{GNF_Sec}).

The following is an interesting observation:
\begin{observ}
Given two permutation braids $u$ and $v$, the decomposition $u \cdot v$ is
{\em left-weighted} if $\partial(u) \wedge v = \varepsilon$ or, equivalently, if
$u v \wedge \Delta = u$. The condition $\partial(u) \wedge v = \varepsilon$ actually means that if we move any crossing from $v$ to $u$, then $u$ will not be anymore a permutation braid.
\end{observ}

\begin{figure}[!htf]
\epsfysize=4cm \centerline{\epsfbox{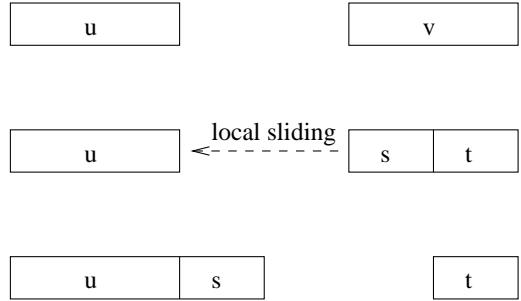}}
\caption{An illustration of a local sliding}\label{local_sliding}
\end{figure}

By this observation, it is easy to give a procedure to find the left-weighted
factorization of the product of two permutation braids $u$ and $v$ as follows. If the decomposition $uv$ is not left-weighted, this means that there is a nontrivial prefix
$s \preceq v$ such that $us$ is still a permutation braid (i.e.
$s \preceq \partial(u))$. The maximal element which satisfies this property is
$s =\partial(u) \wedge v$. Therefore, for transforming the decomposition
$uv$ into a left-weighted one, we have to slide the prefix $s =\partial(u) \wedge v$ from the second factor to the first
one. That is, write $v = st$ and then consider the decomposition $uv = (us)t$, with $us$ as the first
factor and $t$ as the second one. The decomposition $us \cdot t$ is left-weighted by the maximality of $s$.
This action will be called  {\it local sliding} (see Figure \ref{local_sliding}).

Motivated by the idea of local sliding, one wants now to do a cycling in the same manner.
Given a braid in a left normal form $x=\Delta^p x_1 \cdots x_r$, we want now to slide a
part of $x_1$ to $x_r$. This will be done by conjugating a prefix of $\tau^{-p}(x_1)$.
The appropriate prefix is: $\partial(x_r)\wedge \tau^{-p}(x_1)$, which is equal to:
$\iota(x^{-1}) \wedge \iota(x)$. Hence, Gebhardt and Gonz\'alez-Meneses \cite{GeGM2} define:

\begin{defn}
Given $x \in B_n$, define the {\em cyclic sliding ${\mathfrak s}(x)$ of $x$} as the conjugate of $x$ by ${\mathfrak p}(x)=\iota(x^{-1}) \wedge \iota(x)$, that is:
$${\mathfrak s}(x) = {\mathfrak p}(x)^{-1} x {\mathfrak p}(x).$$
\end{defn}

By a series of results, Gebhardt and Gonz\'alez-Meneses \cite[Section 3, Results 3.4-3.10]{GeGM2} show that the cyclic sliding is indeed a generalization of cycling and decycling, and the fact that for every $x \in B_n$, iterated application of cyclic sliding eventually reaches a period, that is, there are integers $N \geq 0$ and $M > 0$ such that
${\mathfrak s}^{M+N}(x) = {\mathfrak s}^N(x)$.

Now, one can define the set of sliding circuits of $x$:
\begin{defn}
An element $y \in B_n$ {\em belongs to a sliding circuit} if ${\mathfrak s}^m(y) = y$ for some $m \geq 1$.

Given $x \in B_n$, {\em the set of sliding circuits of $x$}, denoted by ${\rm SC}(x)$, is the set of all conjugates of $x$ which belong to a sliding circuit.
\end{defn}

Note that ${\rm SC}(x)$ does not depend on $x$ but only on its conjugacy class. Hence, two
elements $x, y \in B_n$ are conjugate if and only if ${\rm SC}(x) = {\rm SC}(y)$.
Therefore, the computation of ${\rm SC}(x)$ and of one element of ${\rm SC}(y)$ will solve the Conjugacy Decision Problem in $B_n$.

The set ${\rm SC}(x)$ is usually much smaller than $\USS(x)$. For example, for
\begin{eqnarray*}
B_{12} \ni x &=& \sigma_7\sigma_8\sigma_7\sigma_6\sigma_5\sigma_4\sigma_9\sigma_8\sigma_7\sigma_6\sigma_5
\sigma_4\sigma_3\sigma_2\sigma_{10}\sigma_9\sigma_8\sigma_7\sigma_6\sigma_5\sigma_4\sigma_3 \cdot \\
& & \cdot \sigma_2\sigma_1\sigma_{11}\sigma_{10}\sigma_9\sigma_8\sigma_7\sigma_6\sigma_5\sigma_4
\sigma_3\sigma_2\sigma_1
\end{eqnarray*}
we have that $|{\rm SC}(x)| = 6$,  but $|\SSS(x)| = |\USS(x)| = 126498$
(see \cite[Section 5]{GeGM2}, based on an example from \cite{Gonz}). On the other hand, the size of the set ${\rm SC}(x)$ still might be exponential in the length of $x$ (for example, if $\delta = \sigma_{n-1} \cdots \sigma_1 \in B_n$, one has $|{\rm SC}(\delta)| = 2^{n-2}- 2$ \cite[Proposition 5.1]{GeGM2}).

Gebhardt and Gonz\'alez-Meneses have proved \cite[Proposition 3.13]{GeGM2} that:
$${\rm SC}(x) = {\rm RSSS}(x)$$
for $x$ satisfying $\ell_s(x)>1$ (where $\ell_s(x) = {\rm sup}_s(x) - {\rm inf}_s(x)$, i.e. the canonical length of elements in the Super Summit Set of $x$), and
$${\rm SC}(x) \subseteq {\rm RSSS}(x)$$
for $x$ satisfying $\ell_s(x)=1$, and in general ${\rm SC} (x)$ is a {\em proper} subset of ${\rm RSSS}(x)$ in this case.

They remark that the case $\ell_s(x) = 1$ in which the sets differ is not irrelevant, since, for example, a periodic braid $x$ which is not conjugate to a power of $\Delta$ has $\ell_s(x) = 1$, but the conjugacy problem for such braids is far from being easy \cite{BGGM3}.

\medskip

As in the previous Summit Sets, the algorithm to solve the CDP/CSP in braid
groups (using sliding circuits) starts by obtaining an element $\hat{x} \in
{\rm SC}(x)$.  We do this as follows: take an element $x$.
Now, apply iterated cyclic sliding on it. Due to the periodic property of
the sliding operation, we will have two integers $m_1,m_2$ ($m_1 < m_2$), which satisfy:
$${\mathfrak s}^{m_1}(x)={\mathfrak s}^{m_2}(x).$$
When having this, the element $\hat{x}={\mathfrak s}^{m_1}(x)$ is
in ${\rm SC}(x)$, since:
$${\mathfrak s}^{m_2-m_1}(\hat{x})=\hat{x}.$$

After finding a representative $\hat{x} \in {\rm SC}(x)$, we have to
explore all the set ${\rm SC}(x)$. This we do in a similar way to the Ultra Summit Set case: There are $\preceq$-minimal elements which conjugate an element
in ${\rm SC}(x)$ to another element there. The number of such possible minimal conjugators
for a given element in ${\rm SC}(x)$ is bounded by the number of Artin generators).
Hence, one can compute the whole ${\rm SC}(x)$ starting by a single
element $\hat{x} \in {\rm SC}(x)$, and then we are done (for more information, see \cite[Section 4.1]{GeGM2} and \cite{GeGM3})

\medskip

Again, as in the previous Summit Sets, the algorithm of Gebhardt and Gonz\'alez-Meneses
\cite{GeGM2} not only computes ${\rm SC}(x)$, but also a graph ${\rm SCG}(x)$,
which determines the conjugating elements. This graph is defined as
follows.

\begin{defn}
Given $x \in B_n$, the directed graph ${\rm SCG}(x)$ is defined by the
following data:
\begin{enumerate}
\item The set of vertices is ${\rm SC}(x)$.

\item For every $y \in {\rm SC}(x)$ and every minimal permutation braid $s$ for
$y$ with respect to ${\rm SC}(x)$, there is an arrow labeled by $s$
going from $y$ to $s^{-1}y s$.
\end{enumerate}
\end{defn}

\medskip

More information about these sorts of Summit sets can be found in the
series of papers \cite{BGGM1,BGGM2,BGGM3} and
\cite{LeeLee3,LeeLee3a,LeeLee4}.

\subsection{An updated summary of the theoretical solution for the conjugacy search problem}

In this section, we give an updated summary for the current status of the complexity of the theoretical solution for the Conjugacy Search Problem. We follow here the nice presentation of Gonz\'alez-Meneses in his talk at Singapore (2007) \cite{GMslides}.

\medskip

As already mentioned, according to Nielsen-Thurston geometric classification (based on \cite{Nielsen} and \cite{Thu88}), there are three types of braids: periodic braids, reducible braids and pseudo-Anosov braids.

A braid $\alpha$ is called {\em periodic} if there exist integers $k,m$ such that $\alpha^k=\Delta^{2m}$. A braid $\alpha$ is called {\em reducible} if it preserves a family of curves, called a {\em reduction system}. A braid is called {\em pseudo-Anosov} if it is neither periodic nor reducible.

\medskip

For the  case of {\bf periodic braids}, Birman, Gebhardt and Gonz\'alez-Meneses \cite{BGGM3} present a polynomial-time algorithm for solving the conjugacy search problem. Almost at the same time, Lee and Lee \cite{LeeLee3b} suggest another entirely different solution for this case.

\medskip

For the case of {\bf reducible braids}, there is a result of Gebhardt and Gonz\'alez-Meneses \cite{GMslides} that these braids fall into exactly two cases:
\begin{enumerate}
\item The braid $\alpha$ is conjugate to a braid with a {\em standard}  reducing curve, which means that the reducing curves are round circles, and hence the Conjugacy Search Problem can be decomposed into smaller problems (inside the tubes). \\
    There is only one problem here: the conjugate braid (with a standard  reducing curve) is in  $\USS(\alpha)$, and for reaching it, one has to make an {\em unknown} number of cycling/decycling (or sliding) steps.
\item The braid $\alpha$ is rigid (i.e. a cycling of the Garside normal form of $\alpha$ is left-weighted as written, or alternatively,  it is a fixed point with respect to cyclic slidings).
\end{enumerate}

For the case of {\bf pseudo-Anosov braids}:  Due to a result of Birman, Gebhardt and Gonz\'alez-Meneses \cite[Corollary 3.24]{BGGM1}, there exists a small power of a pseudo-Anosov braid which is conjugate to a rigid braid. Another result \cite{Me} claims that in the case of pseudo-Anosov braids, the conjugating elements of
the pair $(x,y)$ and the pair $(x^m,y^m)$ coincide, and hence instead of solving the Conjugacy Search Problem in the pair $(x,y)$, one can solve it in the pair $(x^m,y^m)$.
Therefore, one can restrict himself to the case of rigid braids.

\medskip

If we summarize all cases, we get that the main challenges in this direction are:
\begin{enumerate}
\item Solve the Conjugacy Search Problem for rigid braids in polynomial time.
\item Given a braid $x$, find a polynomial bound for the number of cycling/decycling steps one has to perform for reaching an element in $\USS(x)$.
\end{enumerate}

\section{More attacks on the conjugacy search problem}\label{sec_attack_CSP}

There are some more ways to attack the Conjugacy Search Problem,
apart of solving it completely. In this section, we present some
techniques to attack the Conjugacy Search Problem without actually
solving it theoretically.

\subsection{A heuristic algorithm using the Super Summit Sets}
\ \break
Hofheinz and Steinwandt \cite{HS} use a heuristic algorithm for
attacking the Conjugacy Search Problem which is the basis of the
cryptosystems of Anshel-Anshel-Goldfeld \cite{AAFG} and Ko et al.
\cite{KLCHKP}.

Their algorithm is based on the idea that it is
probable that if we start with two elements in the same conjugacy
class, their representatives in the Super Summit Set will not be too
far away, i.e.\ one representative is a conjugation of the other by a
permutation braid.

So, given a pair $(x,x')$  of braids, where $x'=s^{-1}xs$, we do the following steps:
\begin{enumerate}
\item By a variant of cycling (adding a multiplication by $\Delta$ to the first permutation braid, based
 on \cite[Proposition 1]{LeeLee}) and
decycling, we find $\tilde{x} \in \SSS(x)$ and $\tilde{x}' \in
\SSS(x')$.
\item Try to find a permutation braid $P$, such  that $\tilde{x}' = P^{-1} \tilde{x} P$.
\end{enumerate}

In case we find such a permutation braid $P$, since we can follow
after the conjugators in the cycling/decycling process, at the end
of the algorithm we will have at hand the needed conjugator for
breaking the cryptosystem. Note that we do not really need to find
exactly $s$, since each $\tilde{s}$ which satisfies
$x'=\tilde{s}^{-1}x \tilde{s}$ will do the job as well and reveal
the shared secret key.

Their experiments show that they succeed to reveal the shared secret
key in almost $100\%$ of the cases in the Anshel-Anshel-Goldfeld
protocol (where the cryptosystem is based on the Multiple
Simultaneous Conjugacy Problem) and in about $80\%$ of the cases in
the Diffie-Hellman-type protocol.

Note that their attack is special to cryptosystems which are based
on the conjugacy problem, since it depends very much on the fact
that $x$ and $x'$ are conjugate.

\subsection{Reduction of the Conjugacy Search Problem}

Maffre \cite{Maf05,Maf06} presents a deterministic, polynomial
algorithm that reduces the Conjugacy Search Problem in braid group.

The algorithm is based on the decomposition of braids into products
of canonical factors and gives a partial factorization of the
secret: a divisor and a multiple. The tests which were performed on
different keys of existing protocols showed that many protocols in
their current form are broken and that the efficiency of the attack
depends on the random generator used to create the key.

\subsection{Length-based attacks}

A different probabilistic attack on the braid group cryptosystems is
the {\it length-based attack}. In this section, we will sketch its
basic idea, and different variants of this attack on the braid group
cryptosystems. We finish this section with a short discussion about
the applicability of the length-based attack to other groups.

\subsubsection{The basic idea}
The basic idea was introduced by Hughes and Tannenbaum \cite{HT}.

Let $\ell$ be a length function on the braid group $B_n$. In the Conjugacy
Search Problem, we have an instance of $(p,p')$ where $p'=s^{-1}ps$,
and we look for $s$. The idea of a probabilistic length-based attack
to this problem is: if we can write $s=s'\sigma_i$ for a given $i$,
then the length $\ell(\sigma_i s^{-1}ps \sigma_i^{-1})$ should be
strictly smaller than the length $\ell(\sigma_j s^{-1}ps \sigma_j^{-1})$ for $j
\neq i$.

Thus, for using such an attack, one should choose a good length
function on $B_n$ and run it iteratively till we get the correct
conjugator.

\subsubsection{Choosing a length function}\label{length_function}
In \cite{GKTTV06}, we suggest some length functions for this
purposes. The first option is the {\it Garside length}, which is the
length of the Garside normal form by means of Artin generators (i.e.\ if
$w=\Delta_n^r P_1 P_2 \cdots P_k$, then $\ell_{\rm Gar} (w) =
r|\Delta|+|P_1| +|P_2| + \cdots +|P_k|$).

A better length function is the {\it Reduced Garside length} (which
is called {\it Mixed Garside length} in \cite{ECHLPT}). The
motivation for this length function is that a part of the negative
powers of $\Delta_n$ can be canceled with the positive permutation braids.
Hence, it is defined as follows: if $w=\Delta_n^{-r} P_1 P_2 \cdots
P_k$, then:
$$\ell_{\rm RedGar} (w) = \ell_{\rm Gar}(w)- 2 \sum_{i=1}^{\min\{r,k\}}|P_i|.$$
This length function is much more well-behaved, and hence it gives
better performances. But even this length function did not give a
break of the cryptosystems (by the basic length-based attack).

In \cite{HoTs}, Hock and Tsaban checked the corresponding  length
functions for the Birman-Ko-Lee presentation, and they found out
that the reduced length function with respect to the Birman-Ko-Lee presentation
behaves even better than the reduced Garside length function.

\subsubsection{The memory approach}
The main contribution of \cite{GKTTV05} is new improvements to the
length-based attack.

First, it introduces a new approach which uses memory: In the basic
length-based attack, we hold each time only the best conjugator so
far. The problem with this is that sometimes a prefix of the correct
conjugator is not the best conjugator at some iteration and hence it
is thrown out. In such a situation, we just miss the correct
conjugator in the way, and hence the length-based algorithm fails.
Moreover, even if we use a 'look ahead' approach, which means that
instead of adding one generator in each iteration we add several
generators in each iteration, we still get total failure for the
suggested parameters, and some success for small parameters
\cite{GKTTV06}.

In the memory approach, we hold each time a given number (which is
the size of the memory) of possible conjugators which are the best
among all the other conjugators of this length. In the next step, we
add one more generator to all the conjugators in the memory,
and we choose again only the best ones among all the possibilities.
In this approach, in a successful search, we will often have the
correct conjugator in the first place of the memory.

The results of \cite{GKTTV05} show that the length-based attack
with memory is applicable to the cryptosystems of
Anshel-Anshel-Goldfeld and Ko et al, and hence their cryptosystems
are not secure. Moreover, the experiments show that if we increase
the size of the memory, the success rate of the length-based attack
with memory becomes higher.

\subsubsection{A different variant of Length-based attack by Myasnikov and Ushakov}

Recently, Myasnikov and Ushakov \cite{MU07} suggested a different variant of the length-based approach.

They start by mentioning the fact that the {\em geodesic length}, i.e. the length of the shortest path in the corresponding Cayley graph, seems to be the best candidate for a length function in the braid group, but there is no known efficient
algorithm for computing it. Moreover, it was shown by Paterson and Razborov \cite{PR} that the set of geodesic braids in $B_n$ is co-NP complete. On the other hand, many other length functions are bad for the length-based attacks (like the {\em canonical length}, which is the number of permutation braids in the Garside normal form).

As a length function, they choose some approximation function for the geodesic length: they use Dehornoy's handles reduction and conjugations by $\Delta$ (this length function appears in \cite{MSU05,MSU06}). This length function satisfies $|a^{-1}ba| > |b|$ for almost all $a$ and $b$.

Next, they identify a type of braid word, which they call {\em peaks}, which causes problems to the Length-based attacks:
\begin{defn}
Let $G$ be a group, and let $\ell_G$ be a length function on $G$, and
$H = \langle w_1, \dots, w_k \rangle$.
A word $w = w_{i_1} \cdots w_{i_n}$ is called an {\em $n$-peak in $H$ relative to
$\ell_G$} if there is no $1 \leq j \leq n-1$ such that
$$\ell_G(w_{i_1} \cdots w_{i_n}) \geq \ell_G(w_{i_1} \cdots w_{i_j}).$$
\end{defn}

An example of a {\em commutator-type peak} is given in \cite[Example 1]{MU07}:
if $a_1 = \sigma^{-1}_{39} \sigma_{12} \sigma_7 \sigma_3^{-1} \sigma_1^{-1} \sigma_{70} \sigma_{25} \sigma_{24}^{-1}$
and
$a_2 = \sigma_{42} \sigma_{56}^{-1} \sigma_8 \sigma_{18}^{-1} \sigma_{19} \sigma_{73} \sigma_{33}^{-1} \sigma_{22}^{-1}$
then their commutator is a peak:
$a_1^{-1} a_2^{-1} a_1 a_2 = \sigma_7 \sigma_8^{-1}$.

The main idea behind their new variant of the Length-based attack is to add elements from the corresponding subgroup to cut the peaks. By an investigation of the types of peaks, one can see that this is done by adding to the vector of elements all the conjugators and commutators of its elements. By this way, the Length-based attack will be more powerful. For more information and for an exact implementation, see \cite{MU07}.

\subsubsection{Applicability of the length-based approach}

One interesting point about the length-based approach is that it is
applicable not only for the Conjugacy Search Problem, but also for
solving equations in groups. Hence, it is a threat also to the Decomposition
Problem and for the Shifted Conjugacy Problem which was introduced
by Dehornoy (see \cite{Deh06} and Section \ref{shifted} below).

Moreover, the length-based approach is applicable in any group which
has a reasonable length function, e.g.\ the Thompson group, as indeed
has been done by Ruinskiy, Shamir and Tsaban (see \cite{RST06a} and Section
\ref{ThompsonLengthBased} below).

\subsection{Attacks based on linear representations}

A different way to attack these cryptographic schemes is by using
linear representations of the braid groups. The basic idea is to map
the braid groups into groups of matrices, in which the Conjugacy Search
Problem is easy. In this way, we might solve the Conjugacy Search Problem of $B_n$ by lifting the element from the group of matrices back to the braid group $B_n$.

For more information on the linear representations of the braid group, we refer
the reader to the surveys of Birman and Brendle \cite{BiBr} and Paris \cite{Paris}.

\subsubsection{The Burau representations}\label{Burau}
The best known linear representation of the braid group $B_n$ is the
Burau representation \cite{Burau}. We present it here (we partially follow \cite{LeeLee}).

The Burau representation is defined as follows. Let
$\mathbb{Z}[t^{\pm1}]$ be the ring of Laurent polynomials $f(t) =
a_kt^k+a_{k+1}t^{k+1}+ \cdots+ a_mt^m$ with integer coefficients
(and possibly with negative degree terms). Let
${\rm GL}_n(\mathbb{Z}[t^{\pm 1}])$ be the group of $n \times n$
invertible matrices over $\mathbb{Z}[t^{\pm 1}]$. The Burau
representation is a homomorphism $B_n \to {\rm GL}_n(\mathbb{Z}[t^{\pm
1}])$ which sends a generator $\sigma_i\in B_n$ to the matrix:
$$\left( \begin{array}{cccccc}
1 & & & & & \\
 & \ddots & & & & \\
 & & 1-t & t & & \\
 & & 1 & 0 & & \\
 & & & & \ddots & \\
 & & & & & 1
 \end{array} \right) \in {\rm GL}_n(\mathbb{Z}[t^{\pm 1}]),$$
where $1-t$ occurs in row and column $i$ of the matrix.

This representation is reducible, since it can be decomposed into
the trivial representation of dimension $1$ and an irreducible
representation $B_n \to {\rm GL}_{n-1}(\mathbb{Z}[t^{\pm 1}])$ of
dimension $n-1$, called {\it the reduced Burau representation},
which sends a generator $\sigma_i\in B_n$ to the matrix:
$$C_i(t) = \left( \begin{array}{ccccccc}
1 & & & & & \\
 & \ddots & & & & \\
& & 1 & & & &\\
 & & t & -t & 1 & &\\
 & &   &    & 1  & \\
 & & & & & \ddots & \\
 & & & & & & 1
 \end{array} \right) \in {\rm GL}_{n-1}(\mathbb{Z}[t^{\pm 1}]),$$\label{color_Burau_matrix}
where $t$ occurs in row $i$ of the matrix. If $i=1$ or $i=n-1$, the
matrix is truncated accordingly (see \cite{LeeLee}).

Note that these matrices satisfy the braid group's relations:
$$C_i(t)C_j(t) = C_j(t)C_i(t) \mbox{ for } |i -j| > 2$$
$$C_i(t)C_{i+1}(t)C_i(t) = C_{i+1}(t)C_i(t)C_{i+1}(t) \mbox{ for } i = 1, \dots, n-1$$

The Burau representation of $B_n$ is faithful for $n=3$ and it is
known to be unfaithful for $n \geq 5$ (i.e.\ the map from $B_n$ to the
matrices is not injective) \cite{Moody,Moody2,LoPa,Big99}. The case
of $n=4$ remains unknown. In the case of $n \geq 5$, the kernel is
very small \cite{Tu}, and the probability that different braids
admit the same Burau image is negligible.

\medskip

Here is a variant of the Burau representation introduced by Morton \cite{Mor}.
The {\it colored Burau matrix} is a refinement of the Burau matrix by assigning $\sigma_i$ to
$C_i(t_{i+1})$, so that the entries of the resulting matrix have several
variables. This naive construction does not give a group
homomorphism. Thus the induced permutations are considered
simultaneously.  We label the
strands of an $n$-braid by $t_1, \dots , t_n$, putting the label $t_j$ on the
strand which starts from the $j$th point on the right.

Now we define:
\begin{defn} Let $a \in B_n$ be given by
a word $\sigma_{i_1}^{e_1} \cdots \sigma_{i_k}^{e_k}$ , $e_j = \pm 1$.
Let $t_{j_r}$ be the label of the under-crossing strand at the $r$th crossing. Then {\em the colored
Burau matrix} $M_a(t_1, \dots , t_n)$ of $a$ is defined by
$$M_a(t_1, \dots , t_n) = \prod_{r=1}^k (C_{i_r} (t_{j_r} ))^{e_r}.$$
\end{defn}

 The permutation group $S_n$ acts on $\mathbb{Z}[t^{\pm 1}_1 , \dots , t^{\pm 1}_{n-1}]$ from left by
changing variables: for $\alpha\in S_n$, $\alpha(f(t_1, \dots ,
t_n)) = f(t_{\alpha(1)}, \dots , t_{\alpha(n)})$. Then $S_n$ also
acts on the matrix group ${\rm GL}_{n-1}(Z[t^{\pm 1}_1 , \dots , t^{\pm
1}_{n} ])$ entry-wise: for $\alpha\in  S_n$ and $M = (f_{ij} )$,
then $\alpha(M) = (\alpha(f_{ij}))$. Then we have

\begin{defn}
 The {\em colored Burau group} $CB_n$ is:
$$S_n \times GL_{n-1}(\mathbb{Z}[t^{\pm 1}_1
, \dots , t^{\pm 1}_n ])$$ with multiplication
$(\alpha_1,M_1)\cdot(\alpha_2,M_2) = (\alpha_1 \alpha_2,
(\alpha_2^{-1} M_1)M_2)$. The {\em colored Burau representation} $C
: B_n \to CB_n$ is defined by $C(\sigma_i) = ( (i, i + 1),
C_i(t_{i+1}))$.
\end{defn}

It is easy to see the following:
\begin{enumerate}
\item  $CB_n$ is a group, with identity element $(e, I_{n-1})$ and
$(\alpha,M)^{-1} = (\alpha^{-1}, \alpha M^{-1})$,

\item  $C(\sigma_i)$'s satisfy the braid relations and
so $C : B_n \to CB_n$ is a group homomorphism.

\item for $a \in B_n$, $C(a)
= (\pi_a,M_a)$, where $\pi_a$ is the induced permutation and $M_a$ is the colored
Burau matrix.
\end{enumerate}

\medskip

Using the Burau representation, the idea of Hughes \cite{Hu} to
attack the Anshel-Anshel-Goldfeld scheme \cite{AAFG,AAG} is as
follows: take one or several pairs of conjugate braids $(p, p')$
associated with the same conjugating braids. Now, it is easy to
compute their classical Burau image and to solve the Conjugate
Search Problem in the linear group.

In general, this is not enough for solving the Conjugate Search
Problem in $B_n$, because there is no reason for the conjugating
matrix that has been found to belong to the image of the Burau
representation, or that one can find a possible preimage. Since the
kernel of the classical Burau representation is small \cite{Tu}, there is a
non-negligible probability that we will find the correct conjugator
and hence we break the cryptosystem.

\medskip

In a different direction, Lee and Lee \cite{LeeLee} indicate a
weakness in the Anshel-Anshel-Goldfeld protocol in a different
point. Their shared key is the colored Burau representation of a
commutator element.

The motivation for this attack is that despite the change of
variables in the colored Burau matrix by permutations, the matrix in
the final output, which is the shared key, is more manageable than
braids. They show that the security of the key-exchange protocol is
based on the problems of listing all solutions to some Multiple
Simultaneous Conjugacy Problems in a permutation group and in a
matrix group over a finite field. So if both of the two listing
problems are feasible, then we can guess correctly the shared key,
without solving the Multiple Simultaneous Conjugacy Problem in braid
groups.

Note that Lee-Lee attack is special to this protocol, since it uses
the colored Burau representation of a commutator element, instead of
using the element itself. In case we change the representation in
the protocol, this attack is useless.

\subsubsection{The Lawrence-Krammer representation} The
Lawrence-Krammer representation is another linear representation of
$B_n$, which is faithful \cite{Big01,Kr}. It associates with every
braid in $B_n$ a matrix of size ${n \choose 2}$ with entries in a 2-variable Laurent
polynomial ring $\mathbb{Z}[t^{\pm 1}, q^{\pm 1}]$.

Cheon and Jun \cite{CJ} develop an attack against the scheme of
Diffie-Hellman-type protocol based on the Lawrence-Krammer
representation: as in the case of the Burau representation, it is
easy to compute the images of the involved braids in the linear
group and to solve the Conjugacy Problem there, but in general,
there is no way to lift the solution back to the braid groups.

But, since we only have to find a solution to the derived
Diffie-Hellman-like Conjugacy Problem:
\begin{problem}
Given $p, sps^{-1}$ and $rpr^{-1}$, with $r \in LB_n$ and $s \in
UB_n$, find $(rs)p(rs)^{-1}$.
\end{problem}

Taking advantage of the particular form of the Lawrence-Krammer
matrices, which contain many 0's, Cheon and Jun obtain a solution
with a polynomial complexity and they show that, for the parameters
suggested by Ko et al. \cite{KLCHKP}, the procedure is doable, and so the cryptosystem
is not secure.


\section{Newly suggested braid group cryptosystems, their cryptanalysis and their future applications}\label{sec_new_crypto}

In this section, we present recent updates on some problems in the
braid group, on which one can construct a cryptosystem. We also discuss some
newly suggested braid group cryptosystems.

\subsection{Cycling problem as a potential hard problem}
In their fundamental paper, Ko et al. \cite{KLCHKP} suggested some
problems which can be considered as hard problems, on which one can
construct a cryptosystem.  One of the problems is the {\it Cycling
Problem}:
\begin{problem}
Given a braid $y$ and a positive integer $t$ such that $y$ is in the
image of the operator ${\bf c}^t$. Find a braid $x$ such that ${\bf
c}^t(x) = y$.
\end{problem}

Maffre, in his thesis \cite{MafPHD}, shows that the Cycling Problem
for $t = 1$ has a very efficient solution. That is, if $y$ is the
cycling of some braid, then one can find $x$ such that ${\bf c}(x) =
y$ very fast.

Following this result, Gebhardt and Gonz\'ales-Meneses \cite{GeGM} have
shown that the general Cycling Problem has a polynomial solution. The reason
for that is the following result: The cycling operation is
surjective on the braid group \cite{GeGM}. Hence, one can easily
find the $t$th preimage of $y$ under this operation.

\medskip

Note that the decycling operation and cyclic sliding operation are surjective too (the decycling operation is a composition of surjective maps: ${\bf d}(x)=(\tau({\bf c}(x^{-1})))^{-1}$, and the cyclic sliding operation can be written as a composition of a cycling and a decycling \cite[Lemma 3.8]{GeGM2}). Hence,
these problems cannot be considered as hard problems, on which one can construct a cryptosystem \cite{Meneses_PC}.

\medskip

It will be interesting to find new operations on the braid group which their solution can be consider as an hard problem, on which one can construct a cryptosystem.

\subsection{A cryptosystem based on the shortest braid problem}
A different type of problem consists in finding the shortest words
representing a given braid (see Dehornoy \cite[Section
4.5.2]{Deh04}). This problem depends on a given choice of a
distinguished family of generators for $B_n$, e.g., the $\sigma_i$'s
or the band generators of Birman-Ko-Lee.

We consider this problem in $B_{\infty}$ which is the group
generated by an infinite sequences of generators $\{ \sigma_1,
\sigma_2, \dots \}$ subject to the usual braid relations.

The {\it Minimal Length Problem} (or {\it Shortest Word Problem}) is:
\begin{problem}
Starting with a word $w$ in
the $\sigma_i^{\pm 1}$'s, find the shortest word $w'$ which is
equivalent to $w$, i.e., that satisfies $w'\equiv w$.
\end{problem}

This problem is considered to be hard due to the
following result of Paterson and Razborov \cite{PR}:
\begin{prop}
The Minimal Length Problem (in Artin's presentation) is co-NP-complete.
\end{prop}

This suggests introducing new schemes in which the secret key is a
short braid word, and the public key is another longer equivalent
braid word. It must be noted that the NP-hardness result holds in
$B_{\infty}$ only, but it is not known in $B_n$ for fixed $n$.

The advantage of using an NP-complete problem lies in the
possibility of proving that some instances are difficult; however,
from the point of view of cryptography, the problem is not to prove
that some specific instances are difficult (worst-case complexity),
but rather to construct relatively large families of provably
difficult instances in which the keys may be randomly chosen.

Based on some experiments, Dehornoy \cite{Deh04}
suggests that braids of the form
$w(\sigma_1^{e_1} ,\sigma_2^{e_2} , \dots , \sigma_n^{e_n})$ with
$e_i = \pm 1$, i.e., braids in which, for each $i$, at least one of
$\sigma_i$ or $\sigma_i^{-1}$ does not occur, could be relevant.

The possible problem of this approach is that the shortest word
problem in $B_n$ for a fixed $n$ is not so hard. In $B_3$, there is
polynomial-time algorithms for the shortest word problem (see
\cite{Ber} and \cite{Wi} for the presentation by the Artin
generators and \cite{Xu} for the presentation by band generators).
Also, this problem was solved in polynomial time in $B_4$ for the
presentation by the band generators (\cite{KKL} and \cite[Chapter 5]{Lee}).
For small fixed $n$, Wiest \cite{Wi} conjectures for an efficient algorithm for finding
shortest representatives in $B_n$. Also, an unpublished work
\cite{GKTs} indicates that a heuristic algorithm based on a random
walk on the Cayley graph of the braid group might give good results
in solving the Shortest Word Problem.

\medskip
In any case, a further research is needed here in several
directions:
\begin{enumerate}
\item {\bf Cryptosystem direction}: Can one suggest a cryptosystem
based on the shortest word problem in $B_\infty$, for using its hardness due to
Paterson-Razborov?
\item {\bf Cryptanalysis direction}: What is the final status of the shortest word problem in $B_n$
for a fixed $n$?
\item {\bf Cryptanalysis direction}: What is the hardness of the Shortest Word Problem in the Birman-Ko-Lee's presentation?
\end{enumerate}

\subsection{A cryptosystem based on the Shifted Conjugacy Search
Problem}\label{shifted}
Dehornoy \cite{Deh06} has suggested an
authentication scheme which is based on the Shifted Conjugacy
Search Problem.

Before we describe the scheme, let us define the Shifted Conjugacy Search
Problem. Let $x,y \in B_\infty$. We define:
$$x \ast y = x \cdot {\rm d}y \cdot \sigma_1 \cdot {\rm d}x^{-1}$$
where ${\rm d}x$ is the {\it shift} of $x$ in $B_\infty$, i.e.\ ${\rm d}$ is the injective function
on $B_{\infty}$ which sends the generator $\sigma_i$ to the generator $\sigma_{i+1}$ for each $i \geq 1$.
In this context,  the Shifted Conjugacy Search Problem is:
\begin{problem}
Let $s, p \in B_{\infty}$ and $p' = s \ast p$. Find a braid $\tilde{s}$ satisfying
$p' = \tilde{s} \ast p$.
\end{problem}

Now, the suggested scheme is based on the Fiat-Shamir authentication scheme:
We assume that $S$ is a set and $(F_s)_{s\in S}$ is a family of functions of $S$ to itself that
satisfies the following condition:
$$F_r(F_s(p)) = F_{F_r(s)}(F_r(p)),\qquad r,s,p \in S$$
Alice is the prover who wants to convince Bob that she knows the secret key $s$. Then the scheme works as follows:
\begin{protocol}\

{\em Public key:} Two elements $p,p' \in S$ such that $p' = F_s(p)$.

{\em Private keys:} Alice: $s \in S$.

\medskip

{\em Alice:} Chooses a random $r \in S$ and sends Bob $x = F_r(p)$ and $x' = F_r(p')$.

{\em Bob:} Chooses a random bit $c$ and sends it to Alice.

{\em Alice:} If $c=0$, sends $y=r$ (then Bob checks: $x = F_y(p)$ and $x' = F_y(p')$);

If $c=1$, sends $y = F_r(s)$ (then Bob checks: $x' = F_y(x)$).
\end{protocol}

Dehornoy \cite{Deh06} suggests to implement this scheme on Left-Distributive(LD)-systems. A {\it LD-system}
is a set $S$ with a binary operation which satisfies:
$$r \ast (s \ast p) = (r \ast s) \ast (r \ast p).$$

The Fiat-Shamir-type scheme on LD-systems works as follows:
\begin{protocol}\

{\em Public key:} Two elements $p,p' \in S$ such that $p' = s \ast p$.

{\em Private keys:} Alice: $s \in S$.

\medskip

{\em Alice:} Chooses a random $r \in S$ and sends Bob $x = r \ast p$ and $x' = r \ast p'$.

{\em Bob:} Chooses a random bit $c$ and sends it to Alice.

{\em Alice:} If $c=0$, sends $y=r$ (then Bob checks: $x = y \ast p$ and $x' = y \ast p'$);

If $c=1$, sends $y = r \ast s$ (then Bob checks: $x' = y \ast x$).
\end{protocol}

Now, one can use the shifted conjugacy
operation as the $\ast$ operation on $B_{\infty}$ in order to
get a LD-system. So, in this way, one can achieve an authentication scheme
on the braid group with a non-trivial operation \cite{Deh06}.

\begin{remark}
For attacking the Shifted Conjugacy Search Problem, one cannot use the Summit Sets theory,
 since it is not a conjugation problem anymore.
Nevertheless, one still can apply on it the length-based attack, since it is still
an equation with $x$.
\end{remark}

Longrigg and Ushakov \cite{LU07} cryptanalyze the suggestion of Dehornoy, and they show that they can break the scheme (e.g. $24\%$ of success rate for keys of length $100$ in $B_{40}$).  Their idea is that in general cases they can reduce the Shifted Conjugacy Search Problem into the well-studied Conjugacy Search Problem. Based on some simple results, they construct an algorithm for solving the Shifted Conjugacy Search Problem
in two steps:
\begin{enumerate}
\item Find a solution $s' \in B_{n+1}$ for the equation $p' \delta_{n+1}^{-1} = s'd(p) \sigma_1 \delta_{n+1}^{-1}$ in $B_{n+1}$. This part can be done using the relevant Ultra Summit Set.
\item Correct the element $s' \in B_{n+1}$ to obtain a solution $s \in B_n$. This can be done by finding a suitable element $c \in C_{B_{n+1}}(d(p) \sigma_1 \delta_{n+1}^{-1})$ (the centralizer of $d(p) \sigma_1 \delta_{n+1}^{-1}$ in $B_{n+1}$).\\
    The algorithm for computing centralizers presented in \cite{FM02} is based on computing the Super Summit Set, which is hard in general (note that actually the Super Summit Set can be replaced by the Ultra Summit Set and the Sliding Circuits set in Franco and Gonz\'alez-Meneses' algorithm \cite{Meneses_PC}). Hence, Longrigg and Ushakov use some subgroup of the centralizer which is much easier to work with.
\end{enumerate}

In the last part of their paper, they discuss possibilities for hard instances for Dehornoy's scheme, which will resist their attack. Their attack is based on two ingredients:
\begin{enumerate}
\item The Conjugacy Search Problem is easy for the pair
   $$(p' \delta_{n+1}^{-1}, d(p) \sigma_1 \delta_{n+1}^{-1})$$ in $B_{n+1}$.
\item The centralizer
$C_{B_{n+1}}(d(p)\sigma_1 \delta_{n+1}^{-1})$ is "small" (i.e. isomorphic to an Abelian
group of small rank).
\end{enumerate}
Hence, if one can find keys for which one of the properties above is not satisfied, then the attack probably fails.

\medskip

With respect to this scheme, it is interesting to check (see also \cite{Deh06}):
\begin{enumerate}
\item {\bf Cryptanalysis direction:} What is the success rate of a length-based attack on this scheme?
\item {\bf Cryptanalysis direction:} Can one develop a theory for the
Shifted Conjugacy Search Problem which will be
parallel to the Summit Sets theory?
\item {\bf Cryptosystem direction:} Can one suggest a LD-system
on the braid group, which will be secure for the length-based attack?
\item {\bf Cryptosystem direction:} Can one find keys for which the properties above are not satisfied, and for which Longrigg-Ushakov's attack fails?
\item {\bf Cryptosystem direction:} Can one suggest a LD-system on
a different group, which will be secure?
\end{enumerate}

\subsection{Algebraic Eraser}\label{Eraser}
Recently, Anshel, Anshel, Goldfeld and\break Lemieux \cite{AAGL} introduce a new scheme for a cryptosystem which is based on combinatorial group theory.
We will present here the main ideas of the scheme and the potential attacks on it.

\subsubsection{The scheme and the implementation}
We follow the presentation of \cite{KTT}. Let $G$ be a group acting on a monoid $M$ on the left, that is, to each $g \in G$ and each $a \in M$, we associate a unique element denoted $^g a \in M$, such that:
$$ ^1 a = a; \qquad  ^{gh}a = ^g(^ha); \qquad  ^g(ab) = ^ga \cdot ^gb$$
for all $a, b \in M$ and  $g, h \in G$.
The set $M \times G$, with the operation
$(a,g) \circ (b, h)= (a \cdot ^g b, gh)$ is a monoid, which is denoted by $M \rtimes G$.

Let $N$ be a monoid, and $\varphi : M \to N$ a homomorphism. The {\em algebraic eraser}
operation is the function $\star : (N \times G) \times (M \rtimes G) \to (N \times G)$ defined by:
$$(a, g) \star (b, h)= (a \varphi(^g b), gh)$$
The function $\star$ satisfies the following identity:
$$\left( (a, g) \star (b, h) \right) \star (c, r) =
(a, g) \star \left(  (b, h) \circ (c, r) \right)$$
for all $(a, g) \in N \times G$ and $(b,h), (c,r) \in M \rtimes G$.

We say that two submonoids $A,B$ of $M \rtimes G$ are {\em $\star$-commuting} if
$$(\varphi(a), g) \star (b, h)=(\varphi(b), h) \star (a,g)$$
for all $(a, g) \in A$ and  $(b, h) \in B$.
In particular, if $A,B$ $\star$-commute, then:
$\varphi(a)\varphi(^g b) = \varphi(b)\varphi(^h a)$
for all $(a, g) \in A$ and $(b, h) \in B$.

Based on these settings, Anshel, Anshel, Goldfeld and Lemieux suggest the {\em Algebraic Eraser Key Agreement Scheme}. It consists on the following public information:
\begin{enumerate}
\item A positive integer $m$.
\item $\star$-commuting submonoids $A,B$ of $M \rtimes G$, each given in terms of a generating set of size $k$.
\item Elementwise commuting submonoids $C,D$ of $N$.
\end{enumerate}

\medskip

Here is the protocol:
\begin{protocol} \

Alice: Chooses $c \in C$ and $(a_1, g_1),\dots ,(a_m, g_m) \in A$, and sends
$(p, g) = (c, 1) \star (a_1, g_1) \star  \cdots \star (a_m, g_m) \in N \times G$
(where the $\star$-multiplication is carried out from left to right) to Bob.

\medskip

Bob: Chooses $d \in D$ and $(b_1, h_1), \dots , (b_m, h_m) \in B$, and sends
$(q, h) = (d, 1) \star (b_1, h_1) \star \cdots \star (b_m, h_m) \in N \times G$
to Alice.

\medskip

Alice and Bob can compute the shared key:
$$(cq, h) \star (a_1, g_1) \star \cdots \star (a_m, g_m) = (dp, g) \star (b_1, h_1) \star \cdots \star (b_m, h_m)$$
\end{protocol}

For the reason why it is indeed a shared key, see \cite{AAGL} and \cite{KTT}.

Anshel, Anshel, Goldfeld and Lemieux apply their general scheme to a particular case, which they call {\em Colored Burau Key Agreement Protocol} (CBKAP):\\
Fix a positive integers $n$ and $r$, and a prime number $p$.
Let $G = S_n$, the symmetric group on the $n$ symbols $\{ 1, \dots , n\}$. The group
$G=S_n$ acts on $GL_n({\mathbb F}_p(t_1, \dots ,t_n))$ by permuting the variables
$\{t_1, \dots , t_n\}$ (note that in this case the monoid $M$ is in fact a group, and hence, the semi-direct product $M \rtimes G$ also forms a group, with inversion $(a, g)^{-1} = ( ^{g^{-1}} a^{-1}, g^{-1})$ for all $(a, g) \in M \rtimes G$).

Let  $N = GL_n({\mathbb F}_p)$.
The group $M \rtimes S_n$ is the subgroup of\break
$GL_n({\mathbb F}_p(t_1, \dots , t_n)) \rtimes S_n$, generated by $(x_1, s_1),
\dots , (x_{n-1}, s_{n-1})$, where $s_i=(i, i + 1)$, and $x_i =C_i(t_i)$ (see page \pageref{color_Burau_matrix} above), for $i = 2,\dots, n-1$. Recall that the colored Burau group
$M \rtimes G$ is a representation of Artin's braid group $B_n$, determined by mapping each Artin generator $\sigma_i$ to $(x_i, s_i)$, $i = 1, \dots , n-1$.

$\varphi : M \to GL_n({\mathbb F}_p)$ is the evaluation map sending each variable $t_i$ to a fixed element $\tau \in {\mathbb F}_p$.
Let $C = D = {\mathbb F}_p(\kappa)$ is the group of matrices of the form:
$$\ell_1 \kappa^{j_1} + \cdots + \ell_r \kappa^{j_r},$$
with $\kappa \in GL_n({\mathbb F}_p)$ of order $p^n - 1$,
$\ell_1, \dots , \ell_r \in {\mathbb F}_p$, and $j_1, \dots , j_r \in \Z$.

\medskip

Commuting subgroups of $M \rtimes G$ are chosen in a similar way to $LB_n$ and $UB_n$ in Section \ref{DH_braid}. This part is done by a Trusted Third Party (TTP), before the key-exchange protocol starts.

Fix $I_1, I_2 \subseteq \{ 1, \dots , n - 1 \}$ such that for all
$i \in I_1$ and $j \in I_2$, $|i - j|>2$, and
$|I_1|$ and $|I_2|$ are both $\leq n/2$. Then, define
$L = \langle \sigma_i : i \in I_1 \rangle$ and $U = \langle \sigma_j : j \in I_2 \rangle$, subgroups of $B_n$ generated by Artin generators.
From the construction of $I_1$ and $I_2$, $L$ and $U$ commute elementwise. Add to both groups the central element $\Delta^2$ of $B_n$.

Now, they choose a secret random $z \in B_n$. Next, they choose\break $w_1=zw'_1z^{-1}, \dots ,w_k=zw'_kz^{-1} \in zLz^{-1}$ and $v_1=zv'_1z^{-1}, \dots ,v_k=zv'_kz^{-1} \in zUz^{-1}$, each a product of $t$-many
generators. Transform them into Garside's normal form, and remove all
even powers of $\Delta$. Reuse the names $w_1,\dots ,w_k; v_1, \dots, v_k$ for the resulting braids. These braids are made public.

\medskip

Anshel, Anshel, Goldfeld and Lemieux have cryptanalyzed their scheme and the TTP protocol,  and conclude that if the conjugating element $z$ is known, there is a successful linear algebraic attack on CBKAP (see \cite[Section 6]{AAGL}). On the other hand, if $z$ is not known, this attack cannot be implemented. Moreover, they claim that the length-based attack is ineffective against CBKAP because $w_i$ and $v_i$ are not known and for some more reasons.

\subsubsection{The attacks}

There are several attacks on this cryptosystem. Kalka, Teicher and Tsaban \cite{KTT} attack the general scheme and then show that the attack can be applied to CBKAP, the specific implementation of the scheme.

For the general scheme, they show that the secret part of the shared key can be computed (under some assumptions, which also include the assumption that the keys are chosen with standard distributions). They do it in two steps: First they compute $d$ and $\varphi(b)$ up to a scalar, and using that they can compute the secret part of the shared key. They remark that if the keys are chosen by a distribution different from the standard, it is possible that this attack is useless (see \cite[Section 8]{KTT} for a discussion on this point).

In the next part, they show that the assumptions are indeed satisfied for the specific implementation of the scheme. The first two assumptions (that it is possible to generate an element $(\alpha, 1) \in A$ with $\alpha \neq 1$, and that $N$ is a subgroup of $GL_n({\mathbb F})$ for some field ${\mathbb F}$ and some $n$) can be easily checked.
The third assumption (that given an element $g \in \langle s_1,\dots,s_k \rangle$, where
$(a_1, s_1),\dots ,(a_k, s_k) \in M \rtimes G$ are the given generators of $A$, then $g$ can be explicitly expressed as a product of elements of $\{s_1^{\pm}, \dots , s_k^{\pm} \}$), can be reformulated as the {\em Membership Search Problem in generic permutation groups}\;:
\begin{problem}
Given random $s_1,\dots, s_k \in S_n$ and $s \in \langle s_1, \dots , s_k \rangle$,
express $s$ as a short (i.e. of polynomial length) product of elements from
$\{ s_1^{\pm} , \dots , s_k^{\pm} \}$.
\end{problem}

They provide a simple and very efficient heuristic algorithm for solving this problem in generic permutation groups. The algorithm gives expressions of length $O(n^2 \log (n))$, in time $O(n^4 \log (n))$ and space $O(n^2 \log (n))$, and is the first practical one for $n \geq 256$. Hence, the third assumption is satisfied too. So the attack can be applied to the CBKAP implementation.

\medskip

Myasnikov and Ushakov \cite{MU08} attack the scheme of Anshel, Anshel, Goldfeld and Lemieux
from a different direction.
Anshel, Anshel, Goldfeld and Lemieux \cite{AAGL} discuss the security of their scheme and indicate that if the conjugator $z$ generated randomly by the TTP
algorithm is known, then one can attack their scheme by an efficient linear attack,
which can reveal the shared key of the parties. The problem of recovering
the exact $z$ seems like a very difficult mathematical problem since it reduces
to solving the system of equations:
$$\left\{ \begin{array}{c}
w_1=\Delta^{2p_1}zw'_1z^{-1}\\
\vdots \\
w_k=\Delta^{2p_k}zw'_kz^{-1}\\
\ \\
v_1=\Delta^{2r_1}zv'_1z^{-1}\\
\vdots \\
v_k=\Delta^{2r_k}zv'_kz^{-1}\\
\end{array} \right. ,$$
which has too many unknowns, since only the left hand sides are known. Hence, it might be difficult to find the original $z$.

The attack of Myasnikov and Ushakov is a variant of the length-based attack. It is based on the observation that actually any solution $z'$ for the system of equations above can be used in a linear attack on the scheme. Hence, they start by recovering the powers of $\Delta$ which were added, so one can peel the $\Delta^{2p}$ part. In the next step, they succeed in revealing the conjugator $z$ (or any equivalent solution $z'$).

Experimental results with instances of the TTP protocol generated using $|z| = 50$ (which is almost three times greater than the suggested value) showed 100\% success
rate. They indicate that the attack may fail when the length
of $z$ is large relative to the length of $\Delta^2$ (for more details, see \cite[Section 3.4]{MU08}).

\medskip

Chowdhury \cite{Chowd07} shows that the suggested implementation of the Algebraic Eraser scheme to the braid group (the TTP protocol) is actually based on the Multiple Simultaneous Conjugacy Search Problem, and then it can be cracked. He gives some algorithms for attacking the implementation.

\medskip

It will be interesting to continue the research on the Algebraic Eraser key-agreement scheme in several directions:

\begin{enumerate}
\item {\bf Cryptosystem direction}: Can one suggest a different distribution for the choice of keys, so the cryptosystem can resist the attack of Kalka-Teicher-Tsaban?

\item {\bf Cryptosystem direction}: Can one suggest a different implementation (different groups, etc.) for the Algebraic Eraser scheme which can resist the attack of Kalka-Teicher-Tsaban?

\item {\bf Cryptanalysis direction}: Can the {\em usual} length-based approach \cite{GKTTV05} be applied to attack the TTP protocol?

\item {\bf General}: One should perform a rigorous analysis of the algorithm of Kalka-Teicher-Tsaban for the Membership Search Problem in generic permutation groups (see \cite[Section 8]{KTT}).
\end{enumerate}

\subsection{Cryptosystems based on the decomposition problem and the triple decomposition problem}

This section deals with two cryptosystems which are based on different variants of the decomposition problem: Given $a,b=xay\in G$, find $x,y$.

\medskip

Shpilrain and Ushakov \cite{SU06} suggest the following protocol, which is based on the decomposition problem:
\begin{protocol}\

{\em Public key:} $w \in G$.

\medskip

{\em Alice:} chooses an element $a_1 \in G$ of length $\ell$, chooses a subgroup
of the centralizer $C_G(a_1)$, and publishes its generators $A = \{ \alpha_1 , \dots , \alpha_k \}$.

{\em Bob:} chooses an element $b_2 \in G$ of length $\ell$, chooses a subgroup
of $C_G(b_2)$, and publishes its generators $B = \{\beta_1, \dots , \beta_m \}$.

{\em Alice:} chooses a random element $a_2 \in \langle B \rangle$ and sends publicly the
normal form $P_A = N(a_1 w a_2)$ to Bob.

{\em Bob:} chooses a random element $b_1 \in \langle A \rangle$ and sends publicly the
normal form $P_B = N(b_1 w b_2)$ to Alice.

\medskip

{\em Shared secret key:} $K_A = a_1 P_B a_2= b_1 P_A b_2=K_B$.
\end{protocol}

Since $a_1 b_1 = b_1 a_1$ and $a_2 b_2 = b_2 a_2$, we indeed have $K = K_A = K_B$, the shared secret key. Alice can compute $K_A$ and Bob can compute $K_B$.

They suggest the following values of parameters for the protocol: $G = B_{64},\  \ell = 1024$. For computing the centralizers, Alice and Bob should use the algorithm from
\cite{FM02}, but actually they have to compute only some elements from them and not the whole sets.

\medskip

Two key-exchange protocols which are based on a variant of the decomposition problem have been suggested by Kurt
\cite{Kurt}. We describe here the second protocol which is an extension of the protocol of Shpilrain and Ushakov to the {\em triple decomposition problem}:
\begin{problem}
Given $v = x_1^{-1} a_2 x_2$, find $x_1 \in H, a_2 \in A$ and
$x_2 \in H'$ where $H = C_G(g_1, \dots , g_{k_1}), H' = C_G(g'_1, \dots , g'_{k_2})$,
and $A$ is a subgroup of $G$ given by its generators.
\end{problem}

Here is Kurt's second protocol (his first protocol is similar): Let $G$ be a non-commutative monoid with a large number of invertible elements.

\begin{protocol}\

{\em Alice:} picks two invertible elements $x_1, x_2 \in G$, chooses subsets $S_{x_1}\subseteq C_G(x_1)$ and $S_{x_2}\subseteq C_G(x_2)$, and publishes $S_{x_1}$ and $S_{x_2}$.

{\em Bob:} picks two invertible elements $y_1, y_2 \in G$, chooses subsets $S_{y_1} \subseteq C_G(y_1)$ and
$S_{y_2} \subseteq C_G(y_2)$, and publishes $S_{y_1}$ and $S_{y_2}$.

{\em Alice:} chooses random elements $a_1 \in G$, $a_2 \in S_{y_1}$  and $a_3 \in S_{y_2}$ as her private
keys. She sends Bob publicly $(u, v,w)$ where $u = a_1x_1,\ v = x_1^{-1} a_2 x_2,\ w = x_2^{-1} a_3$.

{\em Bob:} chooses random elements $b_1 \in S_{x_1}$, $b_2 \in S_{x_2}$  and $b_3 \in G$ as his private keys.
He sends Alice publicly $(p, q, r)$ where $p = b_1 y_1,\ q = y_1^{-1}  b_2 y_2,\ r = y_2^{-1} b_3$.

\medskip

{\em Shared secret key:} $K=a_1 b_1 a_2 b_2 a_3 b_3$.
\end{protocol}

Indeed, $K$ is a shared key, since Alice can compute $a_1 p a_2 q a_3 r = a_1 b_1 a_2 b_2 a_3 b_3$
and Bob can compute $u b_1 v b_2 w b_3 = a_1 b_1 a_2 b_2 a_3 b_3$.

\medskip

As parameters, Kurt suggests to use $G=B_{100}$ and each secret key should be of length $300$ Artin generators.

\medskip

Chowdhury \cite{Chowd06} attacks the two protocols of Kurt, by observing that by some manipulations one can gather the secret information by solving only the Multiple Simultaneous Conjugacy Search Problem. Hence, the security of Kurt's protocols is based on the solution of the Multiple Simultaneous Conjugacy Search Problem. Since the Multiple Simultaneous Conjugacy Search Problem can be attacked by several methods, Chowdhury has actually shown that Kurt's protocols are not secure.

\medskip

Although Shpilrain and Ushakov indicate that their key-exchange scheme resists length-based attack, it will be interesting to check if this indeed is the situation. Also, it is interesting to check if one can change     the secrets of Kurt's protocols in such a way that it cannot be revealed by just solving the Simultaneous Conjugacy Search Problem. If such a change exists, one should check if the new scheme resists length-based attacks.


\section{Future directions I: Alternate distributions}\label{sec_disctributions}

In this section and in the next section, we discuss some more future directions of research in this area and related areas. This section deals the interesting option of changing the distribution of the generators. In this way, one can increase the security of cryptosystems which are vulnerable when assuming a standard distribution.
In the next section, we deal with some suggestions of cryptosystems which are based on different non-commutative groups, apart from the braid group.

\medskip

For overcoming some of the attacks, one can try to change the
distribution of the generators. For example, one can require that if
the generator $\sigma_i$ appears, then in the next place we give
more probability for the appearance of $\sigma_{i\pm 1}$. In
general, such a situation is called a {\it Markov walk}, i.e.\ the
distribution of the choice of the next generator depends on the
choice of the current chosen generator.

A work in this direction is the paper of Maffre \cite{Maf06}. After
suggesting a deterministic polynomial algorithm that reduces the
Conjugacy Search Problem in braid group (by a partial factorization
of the secret), he proposes a new random generator of keys which is
secure against his attack and the one of Hofheinz and Steinwandt
\cite{HS}.

This situation appears also in the Algebraic Eraser scheme (Section \ref{Eraser}). The attack of Kalka, Teicher and Tsaban \cite{KTT} assumes that the distribution of the generators is standard. They indicate that if the distribution is not standard, it is possible that the attack fails.

\section{Future directions II: Cryptosystems based on different non-commutative groups}\label{sec_diff_groups}

The protocols presented here for the braid groups can be applied to
other non-commutative groups, so the natural question here is:

\begin{problem}
Can one suggest a different non-commutative group where the existing
protocols on the braid group can be applied, and the cryptosystem
will be secure?
\end{problem}

We survey here some suggestions.

\subsection{Thompson group}

When some of the cryptosystems on the braid groups were attacked, it
was natural to look for different groups, with a hope that a similar
cryptosystem on a different group will be more secure and more
successful. The Thompson group is a natural candidate for such a group:
there is a normal form which can computed efficiently, but the
decomposition problem seems difficult. On this base, Shpilrain and
Ushakov \cite{SU05} suggest a cryptosystem.

In this section, we will define the Thompson group, the
Shpilrain-Ushakov cryptosystem, and we discuss its cryptanalysis.

\subsubsection{Definitions and the Shpilrain-Ushakov cryptosystem}
\ \break Thompson's group $F$ is the infinite non-commutative group
defined by the following generators and relations:
$$F = \langle \quad x_0 , x_1 , x_2 , \dots \quad \arrowvert \quad
x_i^{-1} x_k x_i = x_{k+1} \quad (k>i) \quad \rangle$$

Each $w\in F$ admits a unique \emph{normal form} \cite{CFP96}:
$$w = x_{i_1} \cdots x_{i_r}x_{j_t}^{-1} \cdots x_{j_1}^{-1},$$
where $i_1 \le \cdots \le i_r$, $j_1 \le \cdots \le j_t$, and if
$x_{i}$ and $x_{i}^{-1}$ both occur in this form, then either
$x_{i+1}$ or $x_{i+1}^{-1}$ occurs as well. The transformation of an
element of $F$ into its normal form is very efficient \cite{SU05}.

We define here a natural length function on the Thompson group:
\begin{defn}\label{defn:nflen}
\emph{The normal form length} of an element $w \in F$, ${\rm
LNF}(w)$, is the number of generators in its normal form: If
$w=x_{i_1} \cdots x_{i_r}x_{j_t}^{-1} \cdots x_{j_1}^{-1}$ is in
normal form, then ${\rm LNF}(w) = r+t$.
\end{defn}

Shpilrain and Ushakov \cite{SU05} suggest the following key-exchange protocol based on the
Thompson group:
\begin{protocol} \

{\em Public subgroups:} $A,B,W$ of $F$, where $ab=ba$ for all $a \in A$, $b \in B$

{\em Public key:} a braid $w \in W$.

{\em Private keys:} Alice: $a_1 \in A, b_1 \in B$; Bob: $a_2 \in A, b_2 \in B$.

\medskip

{\em Alice:} Sends Bob $u_1 = a_1 w b_1$.

{\em Bob:} Sends Alice $u_2 = b_2 w a_2$

\medskip

{\em Shared secret key:} $K = a_1 b_2 w a_2 b_1$
\end{protocol}

$K$ is a shared key since Alice can compute  $K = a_1 u_2 b_1$ and
Bob can compute  $K = b_2 u_1 a_2$, and both are equal to $K$ since
$a_1,a_2$ commute with $b_1,b_2$.

Here is a suggestion for implementing the cryptosystem \cite{SU05}:
Fix a natural number $s \ge 2$. Let $S_A = \{x_0 x_1^{-1} , \dots ,
x_0 x_s^{-1}\}$, $S_B = \{x_{s+1}, \dots , x_{2s}\}$ and $S_W =
\{x_0, \dots , x_{s+2}\}$. Denote by $A$, $B$ and $W$ the subgroups
of $F$ generated by $S_A$, $S_B$, and $S_W$, respectively. $A$ and
$B$ commute elementwise, as required.

The keys $a_1, a_2\in A$, $b_1, b_2\in B$ and $w\in W$ are all
chosen of normal form length $L$, where $L$ is a fixed integer, as
follows: Let $X$ be $A$, $B$ or $W$. Start with the unit word, and
multiply it on the right by a (uniformly) randomly selected
generator, inverted with probability $\frac 1 2$, from the set
$S_X$. Continue this procedure until the normal form of the word has
length $L$.

For practical implementation of the protocol, it is suggested in
\cite{SU05} to use $s\in\{3,4,\dots,8\}$ and
$L\in\{256,258,\dots,320\}$.

\subsubsection{Length-based attack}\label{ThompsonLengthBased}
We present some attacks on the Ushakov-Shpilrain
cryptosystem.

As mentioned before, the length-based attack is applicable for any
group with a reasonable length function. Ruinskiy, Shamir and Tsaban
\cite{RST06a} applied this attack to the Thompson group.

As before, the basic length-based attack without memory always fails
for the suggested parameters. If we add the memory approach, there
is some improvement: for a memory of size $1024$, there is $11\%$
success. But if the memory is small (up to $64$), even the memory
approach always fails. They suggest that the reason for this
phenomenon (in contrast to a significant success for the
length-based attack with memory on the braid group) is that the
braid group is much closer to the free group than the Thompson
group, which is relatively close to an abelian group.

Their improvement is trying to avoid repetitions. The problem is
that many elements return over and over again, and hence the
algorithm goes into loops which make its way to the solution much
difficult. The solution of this is holding a list of the
already-checked conjugators, and when we generate a new conjugator,
we check in the list if it has already appeared (this part is
implemented by a hash table). In case of appearance, we just ignore
it. This improvement increases significantly the success rate of the
algorithm: instead of $11\%$ for a memory of size $1024$, we now
have $49.8\%$, and instead of $0\%$ for a memory of size $64$, we
now have $24\%$.

In the same paper \cite{RST06a}, they suggest some more improvements
for the length-based algorithm. One of their reasons for continuing
with the improvements is the following interesting fact which was
pointed out by Shpilrain \cite{Shpil}: there is a very simple fix
for key-agreement protocols that are broken in probability less than
$p$: Agree on $k$ independent keys in parallel, and XOR them all to
obtain the shared key. The probability of breaking the shared key is
at most $p^k$, which is much smaller.

\medskip

In a different paper, Ruinskiy, Shamir and Tsaban \cite{SZ06b}
attack the key agreement protocols based on non-commutative groups
from a different direction: by using functions that estimate the
distance of a group element to a given subgroup. It is known that in
general the Membership Problem is hard, but one can use
some heuristic approaches for determining the distance of an element
to a given subgroup, e.g., to count the number of generators which
are not in the subgroup.

They test it against the Shpilrain-Ushakov protocol, which is based
on Thompson's group $F$, and show that it can break about half the
keys within a few seconds.

\subsubsection{Special attack by Matucci}
Some interesting special attack for the Ushakov-Shpilrain
cryptosystem can be found in Kassabov and Matucci \cite{MatKa} and
Mattuci \cite{Mat06}.

\subsection{Polycyclic groups}
Eick and Kahrobaei \cite{EiKa} suggest to use polycyclic groups as the basis of a cryptosystem. These groups are a natural generalization of cyclic groups, but they are
much more complex in their structure than cyclic groups. Hence, their algorithmic theory
is more difficult and thus it seems promising to investigate classes of polycyclic groups as candidates to have a more substantial platform perhaps more secure.

\medskip

Here is one presentation for polycyclic groups:
{\tiny $$\langle a_1, \dots , a_n \ | \ a_i^{-1}a_j a_i = w_{ij} , a_i a_j a_i^{-1}= v_{ij} , a_k^{r_k}= u_{kk},\ {\rm for}\ 1 \leq i < j \leq n , \quad k \in I \rangle$$}
where $I \subseteq \{ 1, \dots , n \}$ and $r_i \in {\mathbb N}$ if $i \in I$ and the right hand sides $w_{ij} , v_{ij} , u_{jj}$ of the relations are words in the generators
$a_{j+1}, \dots , a_n$. Using induction, it is straightforward
to show that every element in the group defined by this presentation can be written in the
form $a_1^{e_1} \cdots  a_n^{e_n}$ with $e_i \in \Z$ and $0 \leq e_i < r_i$ if $i \in I$ (see \cite{sims} for more information).

Eick and Kahrobaei introduce a Diffie-Hellman-type key-exchange which is based on the polycyclic group. As in the braid groups' case,  the cryptosystem
is based on the fact that the word problem can be solved effectively in polycyclic
groups, while the known solutions to the conjugacy problem are far less efficient.
For more information, see \cite{EiKa}.

\medskip

In a different direction, Kahrobaei and Khan \cite{KK} introduce a non-commutative key-exchange scheme which generalizes the classical El-Gamal Cipher \cite{ElGamal}
to polycyclic groups.

\subsection{Miller groups}
Mahalanobis \cite{Mah} suggested some
Diffie-Hellman-type exchange key on Miller Groups \cite{Mil}, which
are groups with an abelian automorphism group.

\subsection{Grigorchuk group}
Garzon and Zalcstein \cite{GaZa} suggest a cryptosystem which is based on the word problem
of the Grigorchuk group \cite{Gri}. Both Petrides \cite{Pet} and Gonz\'alez-Vasco, Hofheinz, Martinez and Steinwandt \cite{GHMS} cryptanalyze this cryptosystem.

The Conjugacy Decision Problem in this group is also polynomial \cite{LMU08}, so this problem cannot be served as a base for a cryptosystem.

\subsection{Twisted conjugacy problem in the semigroup of $2\times 2$ matrices over polynomials}
Shpilrain and Ushakov \cite{SU08} suggest an authentication scheme which is based on the {\em twisted conjugacy search problem}:
\begin{problem}
Given a pair of endomorphisms (i.e., homomorphisms into itself) $\varphi,\psi$
of a group $G$ and a pair of elements $w, t \in G$, find an element $s \in G$
such that $t =  \psi(s^{-1}) w \varphi(s)$ provided at least one such $s$ exists.
\end{problem}

Their suggested platform semigroup $G$ is the semigroup of all $2\times 2$ matrices over
truncated one-variable polynomials over ${\mathbb F}_2$, the field of two elements. For more details, see their paper.

\medskip

\section*{Acknowledgements}

First, I wish to thank the organizers of the PRIMA school and conference on Braids which took place at Singapore in June 2007, Jon Berrick and Fred Cohen, for giving me the opportunity to give tutorial talks
and a conference talk on the fascinating topic of braid group cryptography. Also, I wish to thank the Institute of Mathematical Sciences at the National University of Singapore for hosting my stay.

\medskip

Second, I wish to thank Patrick Dehornoy who has let me use his
survey on braid group cryptography \cite{Deh04}. I have followed his
presentation in many places.

I also want to thank Joan Birman, Rainer Steinwandt and Bert Wiest
for useful communications. I owe special thanks to Juan Gonz\'alez-Meneses and Boaz Tsaban who have helped me in some stages of the preparation of these lecture notes. I want to thank Ben Chacham for some useful corrections.


\end{document}